\documentclass{article}

\usepackage{PRIMEarxiv}

\usepackage[utf8]{inputenc} 
\usepackage[T1]{fontenc}    
\usepackage{hyperref}       
\usepackage{url}            
\usepackage{booktabs}       
\usepackage{amsfonts}       
\usepackage{nicefrac}       
\usepackage{microtype}      
\usepackage{lipsum}
\usepackage{fancyhdr}       
\usepackage{graphicx}       

\usepackage{amsmath}
\usepackage{amsthm}
\usepackage{csquotes}
\usepackage{dirtytalk}
\usepackage[sectionbib, round, authoryear]{natbib}
\usepackage{float}
\usepackage[caption = false]{subfig}
\usepackage{url}

\usepackage[sectionbib]{bibunits}
\defaultbibliographystyle{abbrvnat} 
\defaultbibliography{references}

\newcommand{\Autoref}[1]{%
  \begingroup%
  \def\sectionautorefname{Section}%
   \def\subsectionautorefname{Section}%
  \def\subsubsectionautorefname{Subsubsection}%
  \def\appendixautorefname{Appendix}%
  \def\theoremautorefname{Theorem}%
  \def\remarkautorefname{Remark}%
  \def\corollaryautorefname{Corollary}
  \unskip\autoref{#1}%
  \endgroup%
}

\newcommand{\suppref}[1]{%
  \def\sectionautorefname{Supplementary Material\!\!}
  \autoref{#1}\unskip
}

\newtheorem{theorem}{Theorem}
\newtheorem{remark}{Remark}

\newtheorem{corollary}{Corollary}[theorem]

\newcommand{\hypgeo}[2]{%
  {\vphantom{F}}_{#1}\kern-\scriptspace F_{#2}%
}

\graphicspath{{media/}}     

\title{Control Variate-based Stochastic 
Sampling from the Probability
Simplex
}

\author{
  Francesco Barile \\
  Department of Economics, Management and Statistics \\
  University of Milano-Bicocca \\
  Milan, Italy\\
  \href{mailto:francesco.barile@unimib.it}{\nolinkurl{francesco.barile@unimib.it}}
   \\
   \And
  Christopher Nemeth \\
  School of Mathematical Sciences \\
  Lancaster University \\
  Lancaster, United Kingdom\\
  \href{mailto:c.nemeth@lancaster.ac.uk}{\nolinkurl{c.nemeth@lancaster.ac.uk}} \\
}

\begin{document}
\maketitle

\begin{bibunit}
\begin{abstract}
This paper presents a control variate-based Markov chain Monte Carlo algorithm for efficient sampling from the probability simplex, with a focus on applications in large-scale Bayesian models such as latent Dirichlet allocation. Standard Markov chain Monte Carlo methods, particularly those based on Langevin diffusions, suffer from significant discretization errors near the boundaries of the simplex, which are exacerbated in sparse data settings. To address this issue, we propose an improved approach based on the stochastic Cox--Ingersoll--Ross process, which eliminates discretization errors and enables exact transition densities. Our key contribution is the integration of control variates, which significantly reduces the variance of the stochastic gradient estimator in the Cox--Ingersoll--Ross process, thereby enhancing the accuracy and computational efficiency of the algorithm. We provide a theoretical analysis showing the variance reduction achieved by the control variates approach and demonstrate the practical advantages of our method in data subsampling settings. Empirical results on large datasets show that the proposed method outperforms existing approaches in both accuracy and scalability.
\end{abstract}


\section{Introduction}
There has been increasing interest in probabilistic models where the latent variables, or parameters of interest, are discrete probability distributions over $K$ categories, i.e. vectors lying in the
probability simplex
\begin{align*}
    \Delta_K=\left\{ \left( \omega_1, \omega_2, \ldots, \omega_K \right) : \omega_k \ge 0 , \sum_k \omega_k= 1\right\} \subset \mathbb{R}^K
\end{align*}
Popular examples include topic models, e.g. latent Dirichlet allocation \citep{LDA_Blei}; 
network models, e.g. the stochastic blockmodel \citep{SBM} and, more generally, the mixed membership stochastic blockmodel \citep{MMSBM}. 
Standard approaches to inference over the probability simplex include variational inference \citep{beal2003variational, MAL_001} and Markov chain Monte Carlo schemes, including Gibbs sampling \citep{gilks1995markov}. For the latent Dirichlet allocation model, variational inference \citep{LDA_Blei}, collapsed
variational inference \citep{vb_lda, asuncion2009a} and collapsed Gibbs sampling \citep{Finding_scientific_topics}
methods have been developed.
More recently, online Bayesian variational inference algorithms have been proposed \citep{OVB, OnLearLDA, sparse_stoc_inf_LDA}. 
However, particularly for latent Dirichlet allocation, Markov chain Monte Carlo algorithms have been shown to 
achieve more accurate results faster than variational inference approaches on small to medium corpora \citep{gilks1995markov, vb_lda, asuncion2009a}.
To scale to very large corpora of interest, such as Wikipedia articles, where it is not even feasible to store the whole dataset in computer memory, 
stochastic gradient Markov chain Monte Carlo algorithms for sampling from the probability simplex have been proposed \citep{SGRLD_Patterson_Teh, Baker2018LargeScaleSS}. The most common stochastic gradient sampler is based on the Langevin diffusion \citep{SGLD_Welling_Teh2011}, defined as the solution to the stochastic differential equation
\begin{align*}
\mathrm{d} \theta_t=-\nabla \text{U}\left(  \theta_t \right) \mathrm{d} t+\sqrt{2} \mathrm{d} W_t,
\end{align*}
where $W_t$ is a $K$-dimensional Wiener process. The Langevin diffusion defines a Markov chain whose stationary distribution is $\pi\left( \theta \right) \propto \exp{\{-\text{U}\left(  \theta \right)\}}$,
where $\text{U}\left(  \theta \right)=\sum_{i=1}^N \text{U}_i\left(  \theta \right)$ is the potential function,
with $\text{U}_i\left(  \theta \right)=-\log p\left(x_i \mid \theta\right) - (1/N) \log p(\theta)$ determining the unit's contribution to the Bayesian learning of $\pi\left( \theta \right)$ for data $x_i$ with density $p\left(x_i \mid \theta\right)$ and prior $p(\cdot)$ on $\theta$. However, for general $\pi$, the Langevin dynamics are intractable and samples from its Euler approximation are required, introducing discretization error. In the stochastic gradient setting, the costly
full-data gradient $\nabla \text{U}\left(  \theta_t \right)$ is replaced with an estimator  $\hat{\nabla} \text{U}\left( \theta_t \right)$ calculated on a subsample of the data. This leads to an extra source of error. As a result of these two sources of error, stochastic gradient Markov chain Monte Carlo algorithms target an approximate posterior \citep{SGLD_Welling_Teh2011, teh2016a, Vollmer16}.
 A significant limitation of Langevin-based stochastic gradient Markov chain Monte Carlo methods is that they struggle to sample from constrained spaces. 
\cite{SGRLD_Patterson_Teh} developed the first stochastic gradient Markov chain Monte Carlo method for sampling from the probability simplex,
developing a Riemannian-variant (see \cite{Riemann_manifold_Girolami}) of the stochastic gradient Langevin diffusion \citep{SGLD_Welling_Teh2011, nemeth2021stochastic} to account for the geometry of the simplex space, namely the stochastic gradient Riemannian Langevin dynamics.
Under various parameterizations, \cite{SGRLD_Patterson_Teh} 
find that $\omega_k=|\theta_k|\Large{/}\sum_{k=1}^K |\theta_k|$  performs
the best numerically. However, the authors note that the boundary of the sparse simplex space is where most problems occur using samplers of this type. In the large-scale data setting, the vectors $\omega$ become sparse -i.e., there are many $k$ for which $\omega_k$ is close to zero- pushing them to the boundaries of the simplex.
\cite{Baker2018LargeScaleSS} show that the reason the stochastic gradient Riemannian Langevin dynamics struggle to sample $\omega_k$ when it is near the boundary is due to the bias introduced by the Euler discretization error. 
To counteract this, 
\cite{Baker2018LargeScaleSS} designed a stochastic gradient Markov chain Monte Carlo method based on the Cox--Ingersoll--Ross process \citep{CIR85}, known as the stochastic Cox--Ingersoll--Ross process. This method achieved state-of-the-art
results on the latent Dirichlet allocation model by avoiding the error that is introduced from numerically discretizing a stochastic process. However, 
to scale to large-scale datasets, the stochastic Cox--Ingersoll--Ross process replaces the full-data gradient with an unbiased stochastic approximation. In this paper, we improve on the work of \cite{Baker2018LargeScaleSS} by defining a control variate-based sampling algorithm which significantly reduces the variance in the stochastic gradient and empirically displays a similar level of accuracy as a full-data exact Cox--Ingersoll--Ross process.

\section{Stochastic Cox--Ingersoll--Ross Process}
\label{sec:cir}
The standard Cox--Ingersoll--Ross process \citep{CIR85} with parameters $b$, $a$ and $\sigma$ has the following form
\begin{align}\label{eq:st_cir}
\mathrm{d} \theta_t=b\left(a-\theta_t\right) \mathrm{d} t+ \sigma \sqrt{\theta_t} \mathrm{d} W_t .
\end{align}

The stationary distribution of the diffusion process is  $\operatorname{Gamma}\left(2ab/\sigma^2, 2b/\sigma^2\right)$. Moreover, we
define $\chi^2(\nu, \mu)$ to be the non-central chi-squared distribution with $\nu$ degrees of freedom and non-centrality parameter $\mu$. Then, for $\vartheta_t$ at time $t \geq 0$, the probability distribution of $\theta_{t+h}$ is 
\begin{align}\label{eq:CIR}
    \theta_{t+h} \mid \theta_t=\vartheta_t \sim \frac{(1-e^{-bh})\sigma^2}{4b} W, \quad W \sim \chi^2\left(\frac{4ba}{\sigma^2}, \frac{4b}{\sigma^2}\vartheta_t \frac{e^{-bh}}{1-e^{-bh}}\right),
\end{align}
where $h>0$. 
This transition density allows us to simulate directly from the Cox--Ingersoll--Ross process with no discretization error.  A value of zero for $\theta_{t+h}$ can be achieved if $\sigma^2>2ba$ in 
\eqref{eq:CIR} \citep{CIR85}. This will be particularly useful if the aim is to sample highly sparse $\omega$'s.

A Dirichlet prior $\operatorname{Dir}(\alpha)$ on $\omega$, with density $p(\omega) \propto \prod_{k=1}^K \omega_k^{\alpha_k},$ is a conjugate prior for categorical data $z_i$ of dimension $K$ for $i=1, \ldots, N$, where $z_{i k}$ will be $1$ if data point $i$ belongs to category $k$ and $z_{i j}$ will be zero for all $j \neq k$. This leads to a Dirichlet posterior distribution $\operatorname{Dir}\left(a\right)$, with $a=\alpha+\sum_{i=1}^N z_i$. 
For specific parameterizations, the Cox--Ingersoll--Ross process can be used to sample from a $\operatorname{Gamma}\left(a, 1\right)$ distribution.
Then, using the gamma reparameterization $\omega=\theta / \sum_k \theta_k,$ gives the desired Dirichlet posterior.  \cite{Baker2018LargeScaleSS} set $b = 1$ and $\sigma^2=2$ to avoid identifiability issues.
Apart from some simple cases in which the posterior of $\omega$ can be calculated exactly, 
in many applications, the $z_i$ are latent variables, and they are also simulated as part of a larger Gibbs sampler.
Thus the $z_i$ will change at each iteration of the algorithm. When $N$ is large, standard Markov chain Monte Carlo is prohibitively slow. \cite{Baker2018LargeScaleSS} replace the posterior parameter $a$ with an unbiased estimate using only a subset of $z$ at each iteration: $\hat{a}=\alpha + N / n \sum_{i \in \mathcal{S}} z_i$, where $\mathcal{S} \subset\{1, \ldots, N\}$, with $|\mathcal{S}|=n$. Note that the expectation of the stochastic noise $\xi=\hat{a}-a$ over sampling set $\mathcal{S}$ is $\mathbb{E}_{\mathcal{S}}\left[ \xi \right]=0$ and $\xi$ is typically assumed to be white noise, i.e. $\xi_t$ and $\xi_s$ $(t \neq s)$ are independent. For a Cox--Ingersoll--Ross process as in \eqref{eq:st_cir}, the transformation $h\left( \theta_t \right)=\sigma \sqrt{\theta_t}$ leads to a Langevin diffusion for a generalized gamma distribution. The practical implication of this result 
is that, similar to the stochastic gradient Langevin dynamics, it is possible to replace the Cox--Ingersoll--Ross parameters with an unbiased estimate created from a subsample of data.
Using similar results from the stochastic gradient Langevin dynamics \citep{SGLD_Welling_Teh2011}, it can be shown that the stochastic Cox--Ingersoll--Ross process defines a Markov chain that approximately targets the desired posterior $\operatorname{Gamma}\left(a, 1\right)$.
Replacing the posterior parameter $a$ with a stochastic estimate introduces a source of error. The more accurate this estimator is, the lower the computational cost will be for the same level of accuracy,
and thus it is natural to consider alternatives to the simple estimator $\hat{a}$.

\subsection{Control variate-based stochastic Cox-Ingersoll-Ross algorithm}\label{sec:section4}
The variance of a Monte Carlo estimator can be reduced using control variates \citep{Ripley}, which in our setting involves choosing a set of simple functions $g_i$, $i=1,  \ldots, N$,  which we refer to as \textit{control variates}, and whose sum, $\sum_{i=1}^N g_i\left( \theta \right)$, can be evaluated for any $\theta$. 
We can obtain an unbiased estimator of the full gradient as
 \begin{align}\label{eq:cv_est}
     \hat{\nabla} \text{U}_{\text{CV}} \left(\theta\right)=
     \sum_{i=1}^N g_i \left( \theta \right) + \frac{N}{n} \sum_{i \in \mathcal{S}} \left( \nabla \text{U}_i\left(  \theta \right) -g_i\left( \theta \right)  \right)
 \end{align}
where $\mathcal{S}$ is a random sample, without replacement, from $\{1, \ldots, N\}$. 
One approach to choosing the control variate function $g_i\left( \theta \right)$ that is often used in practice, is to (i) use stochastic gradient descent to find an approximation to the mode of the distribution $\pi$, which we denote as $\hat{\theta}$; and (ii) set $g_i\left( \theta \right)=\nabla \text{U}_i(\hat{\theta})$. The intuition behind this idea is that if each $g_i\left( \theta \right) \approx \nabla \text{U}_i(\theta)$, then this
estimator can have a much smaller variance than the simple gradient estimator, where $g_i(\theta)=0$ $\forall i$.
Specifically, it has been shown \citep[e.g. Chapter 3 of ][]{fearnhead2024scalable} that in the large data setting, and under certain Lipschitz assumptions on the gradient, the variance of the simple gradient estimator $\hat{\nabla} \text{U}(\theta)$ scales as $O(N^2/n)$, whereas the variance of the control variate-based estimator \eqref{eq:cv_est} scales as $O(N/n)$. Therefore, for the same level of accuracy, we can reduce the computational cost by $O(N)$ if we use control variate-based gradient estimators.

We now apply this control variate idea 
to the stochastic Cox-Ingersoll-Ross process \citep{Baker2018LargeScaleSS}.
For each $k=1, \ldots, K$, let $a_k=\alpha_k+\sum_{i=1}^N z_{i k}$. The posterior of interest can be cast as a Generalized Multivariate Gamma distribution \citep{gmg_book}, $\theta \mid z \sim \text{GMG}\left( a_1, \ldots, a_K; 1; 0 \right)$ and using the conditional independence property  $f(\mathbf{\theta} \mid z)=\prod_{k=1}^K f(\theta_k \mid z)$, the marginal posterior is $\theta_k \mid z \stackrel{\text {ind}}{\sim} \operatorname{Gamma}\left(a_k, 1\right)$, with associated \textit{pdf} $f(\theta_k \mid z)\propto\theta_k^{a_k-1} e^{-\theta_k}$.
Thus, we can define the control variate-based gradient estimator for each categorical variable independently. The posterior mode is $\hat{\theta}_k=\text{Mode}(\theta_k \mid z)=a_k-1$. Since mode is known exactly, step (i) is cost-free and $\sum_{i=1}^N g_i \left( \theta \right)=0$. 
The $i$th component of the posterior gradient is
\begin{align}\label{eq:grad_cv_supp}
 \text{U}_i^\prime\left(  \theta_k \right)=-\frac{\partial}{\partial \theta_k} \log p\left(z_{ik} \mid \theta_k \right)
- (1/N) \frac{\partial}{\partial \theta_k} \log p\left( \theta_k \right)
=-\frac{1}{\theta_k}\left( \frac{\alpha_k-1}{N} + z_{i k} \right)-\frac{1}{N}.  
\end{align}
Trivial calculation leads 
\eqref{eq:cv_est}
to 
\begin{align*}
    \hat{U}^\prime_{\text{CV}} \left(\theta_k\right) =  -\frac{\partial}{\partial \theta_k} \log \pi \left( \theta_k \right) = -\frac{\hat{a}_k-1}{\theta_k} + \frac{\hat{a}_k-1}{a_k-1}.
\end{align*}
Solving for $\pi \left(\theta_k \right)$ 
\begin{align}\label{eq:scir_cv_pd}
\theta_k \mid z \stackrel{\text {ind}}{\sim} \operatorname{Gamma}(\hat{a}_k, \hat{b}_k),  \quad  \hat{a}_k=\alpha_k + N / n \sum_{i \in \mathcal{S}} z_{ik}, \; \; \hat{b}_k=\frac{\hat{a}_k-1}{a_k-1}
\end{align}
where $\mathcal{S} \subset\{1, \ldots, N\}$, with $|\mathcal{S}|=n$. \autoref{eq:scir_cv_pd} represents the stationary distribution of a 
Cox--Ingersoll--Ross process with parameters 
 $a=\frac{\hat{a}_k}{\hat{b}_k}$, $b=1$ and $\sigma^2=\frac{2}{\hat{b}_k}$ targeting the posterior distribution $\operatorname{Gamma}\left(a_k, 1\right),$ with variance-reduced stochastic gradients. 
The differential form of the proposed control variate-based stochastic Cox--Ingersoll--Ross process is:
\begin{align}\label{eq:scir_cv}
\mathrm{d} \theta_t=\left(\frac{\hat{a}}{\hat{b}}-\theta_t\right) \mathrm{d}  t+ \sqrt{ \frac{2}{\hat{b}} \theta_t} \mathrm{d}  W_t,
\end{align}
where the subscript $k$ is omitted here to ease the notation.
If at iteration $m \geq 0$ we have state $\vartheta_m$, then its transition density is:
\begin{align}\label{eq:scir_cv_trans}
\hat{\theta}_{m+1} \mid \hat{\theta}_m=\vartheta_m \sim \frac{1-e^{-h}}{2\hat{b}_m} W, \quad W \sim \chi^2\left(2 \hat{a}_m, 2 \vartheta_m \hat{b}_m \frac{ e^{-h}}{1-e^{-h}}\right)
\end{align}
The stepsize $h > 0$ determines how often $\hat{a}_m$ is resampled in a unit time interval rather than the granularity of the discretization. 
A value of zero for the process in \eqref{eq:scir_cv} is reached if
$\hat{a}<1$, which, in turn, is satisfied if $\alpha<1$ in a completely sparse setting. On the other hand, this can lead to $\hat{b}<0$ in a highly sparse setting. This is not allowed since $\hat{b}$ appears in the variance term of 
\eqref{eq:scir_cv}. One solution to this is to use a stratified scheme to sample $\hat{a}$. An alternative parametrization to \eqref{eq:scir_cv} that allows for negative values of $\hat{b}$ is provided in \autoref{app:app3} of the \suppref{sec:supp_mat}. However, its theoretical results (which are given in the \suppref{sec:supp_mat}) are not as easy to interpret as those for \eqref{eq:scir_cv}, which are presented in \Autoref{subsec:Theo_analysis}.

\begin{figure}[t!]
\centering
\includegraphics[width = 0.8\textwidth,
height=7cm
]
{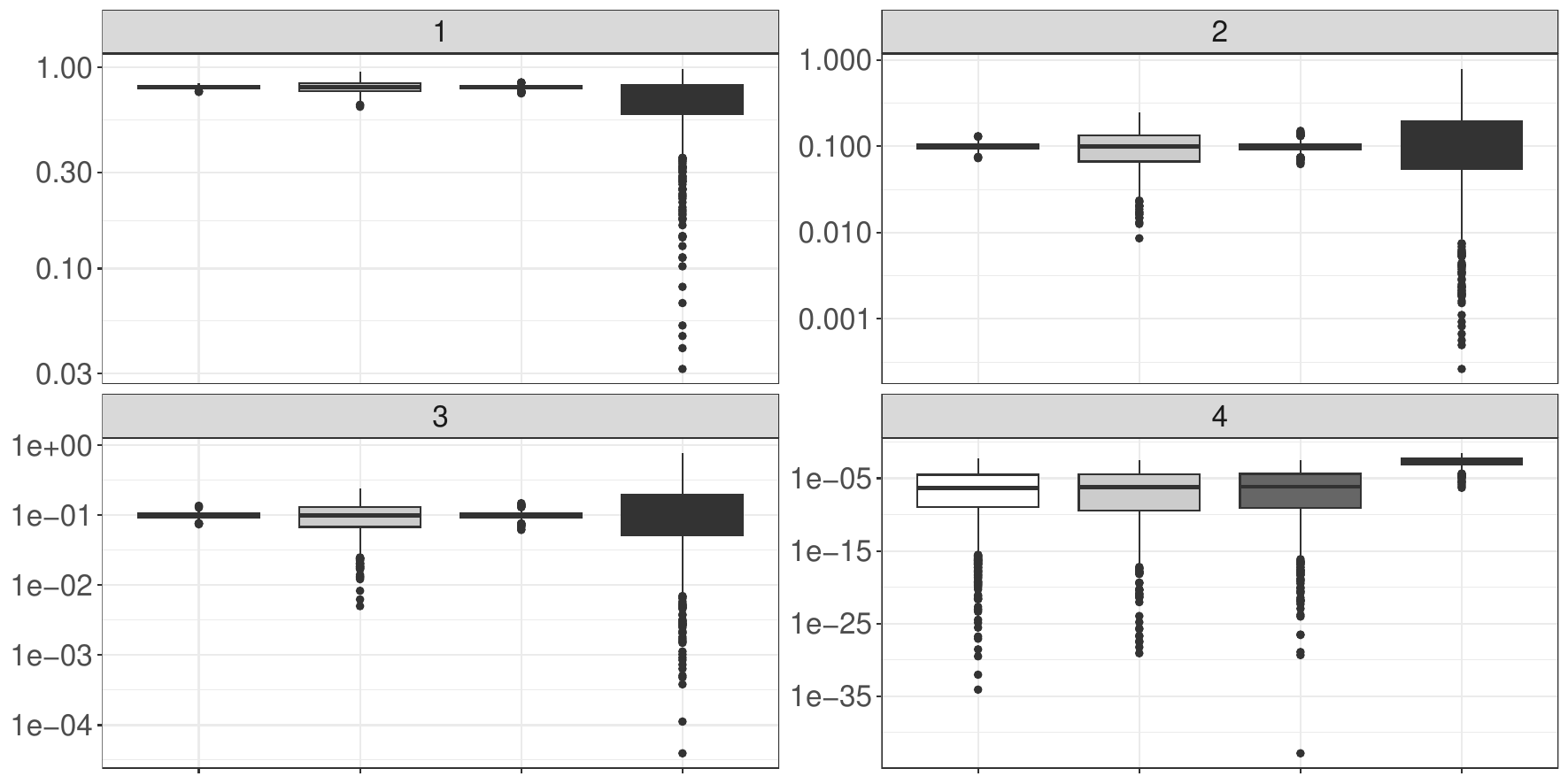}
\caption{Boxplots of a sample from
(from left to right): Exact posterior (white),  the stochastic Cox--Ingersoll--Ross process of \cite{Baker2018LargeScaleSS} (light gray), the control variate-based stochastic Cox--Ingersoll--Ross process (dark gray) and the stochastic gradient Riemannian Langevin dynamics of \cite{SGRLD_Patterson_Teh} (extra-dark gray). 
}\label{fig:run_ex_1}
\end{figure}

In order to demonstrate the improvements achieved using control variates over the simple estimator in the stochastic Cox--Ingersoll--Ross process, we provide an experiment similar to the example in \cite{Baker2018LargeScaleSS}. We simulate from a sparse simplex parameter $\omega$ of dimension $K=10$ with $N=1000$. We set $\sum_{i=1}^N z_{i 1}=800$, $\sum_{i=1}^N z_{i 2}=\sum_{i=1}^N z_{i 3}=100$, and $\sum_{i=1}^N z_{i k}=0$, for $3 < k \leq 10$. The prior parameter $\alpha$ was set to $0.1$ for all components, leading to a highly sparse Dirichlet posterior.
\autoref{fig:run_ex_1} provides boxplots from a sample of the first four components of $\omega$
based on $1\,000$
iterations after a burn-in of $1\,000$.
We set the subsample size $n=10$ and $h=0.5$. 
This experiment highlights the benefit of using a control variate framework. 
It is worth noting that in a completely sparse setting, i.e. $\sum_{i=1}^N z_{i k}=0$, 
the Gamma posterior mode equals $\hat{\theta}_k=\text{Mode}(\theta_k \mid z)=0$
for a prior hyperparameter $\alpha<1$. In this case, the control-variate framework naturally adapts to this setting and takes no action, as illustrated in the fourth panel of \autoref{fig:run_ex_1}. \\

\subsection{Theoretical analysis}\label{subsec:Theo_analysis}
In the following theoretical analysis, we aim to target a $\operatorname{Gamma}(a, 1)$ distribution. We use the control variate-based stochastic Cox--Ingersoll--Ross algorithm \eqref{eq:scir_cv_trans}, with fixed stepsize $h>0$ and $M$ iterations, to yield samples $\hat{\theta}_m$ for $m=1, \ldots, M$. \autoref{theo:theorem_mgf} gives the moment-generating function of $\hat{\theta}_m$. 
\begin{theorem}\label{theo:theorem_mgf}
Let $\hat{\theta}_M$ be generated by the control variate-based stochastic Cox--Ingersoll--Ross process defined in \eqref{eq:scir_cv_trans} starting from $\theta_0$ after $M$ steps with stepsize $h$. The moment-generating function of $\hat{\theta}_M$ is 
\begin{align*}
\mathcal{M}_{\hat{\theta}_M}(s) & = MGF_{\hat{\theta}_M \mid \theta_{0}}(s) =  \exp{\left\{\theta_{0} r^{(1:M)}(s)\right\} }C_M(s) \prod_{m=1}^{M-1}  C^{(m:M)}(s)
\end{align*}
where
\begin{align*}
r^{(m:M)}(s)= \frac{s  e^{-(M-m+1)h}  
}{
D^{\binom{M-m+1}{M-m}}(s) },  
\quad 
C_M(s)=\left( \frac{ \hat{b}_M-s(1-e^{-h })  }{\hat{b}_M} \right)^{-\hat{a}_M}, 
\quad 
C^{(m:M)}(s)= \left( \frac{ 
D^{\binom{M-m+1}{M-m}}(s)
 }
{ D^{\binom{M-m+1}{M-m}-1}(s) 
} \right)^{-\hat{a}_m}
\end{align*}
and we define $D^R(s) = 1  - s \left(1-e^{-h} \right)
 \sum_{l=1}^{R} \frac{e^{-(l-1)h}}{\hat{b}_{M-(l-1)}}$ for ease of notation.
\end{theorem}

The proof of this result follows by recursively applying properties of the non-central chi-squared distribution and it is provided in the \autoref{app:app1} of the \suppref{sec:supp_mat}.
Deriving the moment-generating function enables us to find the non-asymptotic bias and variance of the control variate-based stochastic Cox--Ingersoll--Ross process, which are more interpretable than
the moment-generating function itself. 

\begin{corollary}\label{theo:coro_mgf}
Given the moment-generating function of \autoref{theo:theorem_mgf}, it follows that 
\begin{align}\label{eq:expected_value}
        \mathbb{E}\left[\hat{\theta}_M  \right] 
 & =  
 \theta_0 e^{-Mh} + (1-e^{-Mh} ) \left(a-1 \right)  \mathbb{E}\left[\frac{\hat{a}}{\hat{a}-1}   \right],
\end{align}
\begin{align}\label{eq:variance}
    \mathbb{V}ar\left[\hat{\theta}_M  
    \right]
    & = 
2 \left( a-1\right) \mathbb{E}\left[ \frac{1  }{\hat{a}-1 } \right] \left\{ \theta_0 \mathcal{C}_1 + \mathcal{C}_3  \left( a-1\right) \mathbb{E}\left[ \frac{\hat{a}  }{\hat{a}-1 } \right]\right\}
 +  \mathcal{C}_2  \left( a-1\right)^2 \mathbb{E}\left[ \frac{\hat{a}  }{\left(\hat{a}-1\right)^2 } \right] 
\end{align}
where $\mathcal{C}_1$, $\mathcal{C}_2$ and $\mathcal{C}_3$ are constants depending on $h$ and $M$ . Specifically:
\begin{align*}
& \mathcal{C}_1=( e^{-Mh} - e^{-2Mh} ), \quad
     \mathcal{C}_2= ( 1- e^{- h }  )^2 
     \left( \frac{1-e^{-2Mh}}{1-e^{-2h}} \right), \\
    &  \mathcal{C}_3=( 1- e^{- h }  )^2 \left[ \frac{e^{-Mh}-e^{-h} }{\left( 1-e^{-h}\right) \left( e^{-h}-1\right) } - \frac{e^{-2Mh}-e^{-2h}}{\left( 1-e^{-h}\right) \left( e^{-2h}-1\right)} \right]
\end{align*}
\end{corollary}
The expectation $\mathbb{E}\left[\hat{\theta}_M  \right] $ and variance $\mathbb{V}ar\left[\hat{\theta}_M  
\right]$ depend on the distribution
of the random generating mechanism of the mini-batch estimator $\hat{a}$ through the expectation of non-linear functions we define as $\phi\left( \cdot \right)$.
Similar findings are provided in \autoref{app:app3} of the \suppref{sec:supp_mat} for the alternative parametrization.
Let 
$\sigma^2(a)= (a-1)^{-2} \mathbb{V}\text{ar}\left[\hat{a}  \right]$. 
Expanding expectations of the type  $\mathbb{E}\left[\phi\left( \hat{a} \right)\right]$ at $\mathbb{E}\left[ \hat{a}\right]=a$ and assuming the first two moments of $\hat{a}$ appropriately summarize its distributional properties, it follows that
\begin{align}\label{eq:mean_approx}
    \mathbb{E}\left[\hat{\theta}_M  \right]  \approx \mathbb{E}\left[ \theta_M \right] + (1-e^{-Mh}) \, \sigma^2(a),
\end{align}
\begin{align}\label{eq:var_approx}
\mathbb{V}ar\left[\hat{\theta}_M  \right]    \approx \mathbb{V}ar\left[\theta_M  \right] + \mathcal{B}_1 \sigma^2(a) + \mathcal{B}_2 \left[\sigma^2(a)\right]^2, 
\end{align}
where
$\mathcal{B}_1= \mathbb{V}ar\left[\theta_M  \right] +( 1-e^{-Mh})^2 + \mathcal{C}_2$, \, $\mathcal{B}_2=2\mathcal{C}_3$
and
\begin{align}\label{eq:exact_cir_moments}
 \mathbb{E}\left[\theta_M  \right] =\theta_0 e^{-Mh} + a \, (1-e^{-Mh} ), \quad \mathbb{V}\text{ar}\left[\theta_M  \right]  = 2 \theta_0 \, ( e^{-Mh} - e^{-2Mh} ) + a \, ( 1-e^{-Mh})^2. 
\end{align} 
Thus,
the non-asymptotic expectation and variance result approximatively in simple sums of the expectation $ \mathbb{E}\left[\theta_M  \right]$ and variance $\mathbb{V}\text{ar}\left[\theta_M  \right]$ of the exact underlying Cox--Ingersoll--Ross process, respectively, and a quantity involving the variance of the estimate $\hat{a}$. This is of similar finding to the variance of the stochastic Cox--Ingersoll--Ross process \citep{Baker2018LargeScaleSS}.\\
Since $\mathbb{E}_\pi\left[\theta\right]=a$, 
    $\lim_{M \rightarrow \infty}  \Big| \frac{1}{M}\sum_{m=1}^M \mathbb{E}\left[\hat{\theta}_m\right]- \mathbb{E}_\pi\left[\theta\right] \Big| \approx 
    \sigma^2(a)$.
This result shows that, while the accuracy of the process can be improved for $M$ sufficiently large, there is a systematic component due to the stochastic noise intrinsic in the estimate $\hat{a}$ and the overall asymptotic accuracy is inflated relative to the exact underlying Cox--Ingersoll--Ross process.
Nevertheless, the extra term is substantially lower than that of the stochastic Cox--Ingersoll--Ross process of \cite{Baker2018LargeScaleSS} and the bias term $\sigma^2(a)$ in the estimator can be removed in a post-hoc correction by estimating $\mathbb{V}\text{ar}\left[\hat{a}\right]$ once using every $\ell$ iterations a larger set of observations. We also note that compared to \cite{Baker2018LargeScaleSS} our approach further benefits from non-sparse settings, as evident as evident from the expression of $\sigma^2(a)$ and seen in \autoref{fig:run_ex_1}.
We investigate this in the
\autoref{app:app3} of the \suppref{sec:supp_mat} where
\autoref{fig:var_comparison} illustrates the gain achieved by the control variate-based stochastic Cox--Ingersoll--Ross process over the stochastic Cox--Ingersoll--Ross process of \cite{Baker2018LargeScaleSS} in terms of variance reduction and further illustrates a comparison between the variance of the control variate-based stochastic Cox--Ingersoll--Ross process under our two proposed parametrizations. We note that the alternative parametrization is highly accurate and empirically displays an almost identical level of accuracy as a full-data exact Cox--Ingersoll--Ross process. 

\section{Real-World Application}\label{sec:section5}
\begin{figure}[t!]
\centering
\includegraphics[
height=7cm,
width = 0.8\textwidth] {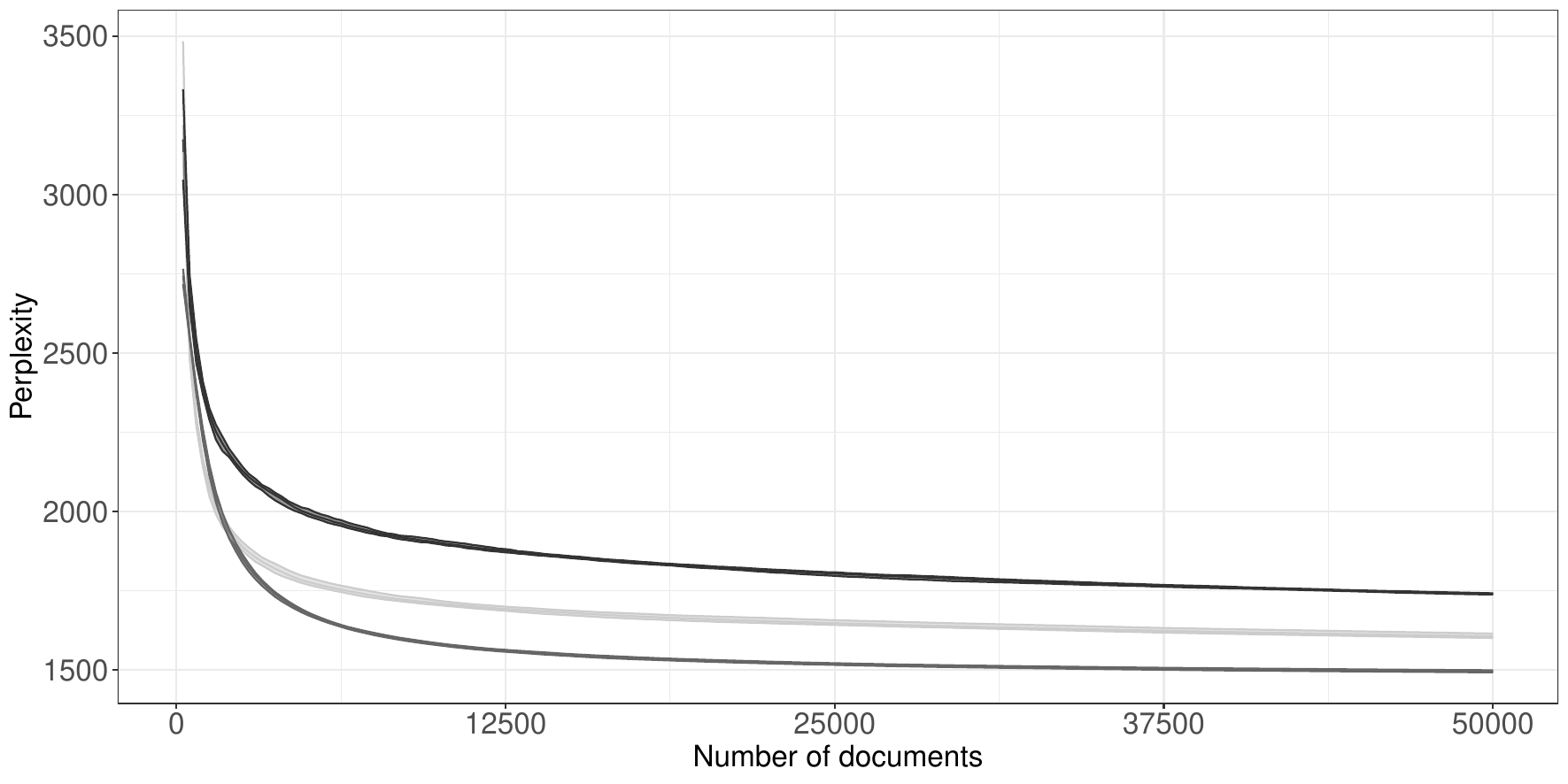}
\caption{The perplexity of the control variate-based stochastic Cox--Ingersoll--Ross process (dark gray), the stochastic Cox--Ingersoll--Ross process of \cite{Baker2018LargeScaleSS} (light gray) and the stochastic gradient Riemannian Langevin dynamics of \cite{SGRLD_Patterson_Teh} (extra-dark gray) when used to sample from the latent Dirichlet allocation model applied to Wikipedia documents.}\label{fig:lda_perp}
\end{figure}

In this section, we empirically compare the control variate-based stochastic Cox--Ingersoll--Ross process to 
its competitors
on the challenging latent Dirichlet allocation \citep{LDA_Blei} model.
The model consists of
$K$ topics each with its distribution $\omega_k$ over the words in the vocabulary drawn from a symmetric Dirichlet prior with hyper-parameter $\beta$. A document $w_d$ is modelled as a mixture of topics.
The model is a
generative process where documents are produced as a set of words by drawing a topic assignment $z_{di}\stackrel{\text {iid}}{\sim} \eta_d$ for each word $w_{di}$ in document $w_d$ and then drawing the word from the corresponding topic $\omega_{z_{di}}$. The latent Dirichlet allocation model is a good test case for the control variate-based stochastic Cox--Ingersoll--Ross process as the computational cost for this model can be substantially reduced since $\omega_k$ is usually high-dimensional for real-world vocabulary sizes.
Performance is evaluated by measuring the predictive ability of the trained model on a held-out test set. 
A metric frequently used for this purpose is perplexity.
\autoref{fig:lda_perp} shows the perplexity for
the latent Dirichlet allocation model applied to a dataset of scraped Wikipedia documents.
The vocabulary used 
results in a 
size of approximately $8\,000$ words. 
At each iteration, subsamples
of $50$ documents are used. The perplexities were estimated
on a separate holdout set of $1\,000$ documents, split $90/10$ training/test for five runs using different seeds, which highlights variability. Similar to \cite{Baker2018LargeScaleSS, SGRLD_Patterson_Teh}, for all methods we use a decreasing stepsize
scheme of the form $h_m=h\left[ 1+m/\tau\right]^{-\kappa}$.
This experiment illustrates the improvements given by the control variate-based stochastic Cox--Ingersoll--Ross process over the stochastic gradient Riemannian Langevin dynamics \citep{SGRLD_Patterson_Teh} in removing the discretization error and improvements over the stochastic Cox--Ingersoll--Ross process \citep{Baker2018LargeScaleSS} in reducing the variance of the unbiased gradient estimator.

\end{bibunit}

\clearpage

\begin{bibunit}

\bigskip
\section*{}\label{sec:supp_mat}
\begin{center}
{\Large\bf Supplementary Material for \say{Control Variate-based Stochastic 
Sampling from the Probability
Simplex} }
\end{center}

The \suppref{sec:supp_mat} is organized as follows. \Autoref{app:app1} includes proofs of the theoretical results stated in \Autoref{subsec:Theo_analysis}. 
An alternative parametrization that equivalently targets approximately the desired posterior is presented in the \Autoref{app:app3}. Further details on the latent Dirichlet allocation experiment are also provided in the \Autoref{app:app4}.

\allowdisplaybreaks

\appendix
\section{Appendix 1}\label{app:app1}
Here, we provide proofs of \autoref{theo:theorem_mgf} and \Autoref{theo:coro_mgf}.
\subsection*{Proofs of theoretical results}
\begin{proof}[Proof of \autoref{theo:theorem_mgf}]\label{proof_theo1}
First, for any step $m=1, 2, \ldots, M$, let us define the following quantities  
\begin{align}\label{eq:supp_eq}
r_m(s)= \frac{s \hat{b}_m e^{-h}}{\hat{b}_m-s(1-e^{- h})}, \qquad C_m(s)=\left( \frac{ \hat{b}_m-s(1-e^{- h})  }{\hat{b}_m} \right)^{-\hat{a}_m}. 
\end{align}
Moreover, define 
$$r^{(m:M)}(s)=r_m \circ r_{m+1}\circ \ldots \circ r_{M-1} \circ  r_{M}(s),  
$$
and, for $m=1, \ldots, M-1$, 
$$ C^{(m:M)}(s)= C_{m}( r^{((m+1):M)}(s)).$$

Suppose that $\hat{\theta}_1 \mid \theta_0$ is a control variate-based stochastic 
Cox--Ingersoll--Ross process, starting at $\theta_0$ and running for time $h$. Then we can immediately write down the moment-generating function of $\hat{\theta}_1$, $M_{\hat{\theta}_1}(s)$, using the moment-generating function of the non-central chi-squared distribution
\begin{align*}
\mathcal{M}_{\hat{\theta}_1}(s)= 
\mathbb{E}[ e^{s \hat{\theta}_1} \mid \theta_0] & =
\left[ 
\frac{\hat{b}_1 - s ( 1- e^{- h} )}{\hat{b}_1}
\right]^{-\hat{a}_1}
\exp{ \left\{ \theta_0\frac{s  \hat{b}_1 e^{- h} }{\hat{b}_1 - s ( 1- e^{- h} )  } \right\}
}  \\
& =C_1(s) \exp{\left\{\theta_{0} r_1(s)\right\} }
\end{align*}
We can use this to find $\mathbb{E}[ e^{s \hat{\theta}_M} \mid \hat{\theta}_{M-1}]$, and then take expectation of this with respect to $\hat{\theta}_{M-2}$ to find $\mathbb{E}[ e^{s \hat{\theta}_M} \mid \hat{\theta}_{M-2}]$. This is possible because $\mathbb{E}\left[ \mathbb{E}[ e^{s \hat{\theta}_M} \mid \hat{\theta}_{M-1}] \mid \hat{\theta}_{M-2} \right]$ has the form 
$C_M(s) C^{\left( (M-1):M \right)}(s) \exp{\left\{\hat{\theta}_{M_1} r^{\left((M-1):M\right)}(s)\right\}}$ where $C_M(s)$, $C^{\left( (M-1):M \right)}(s)$ and $ r^{\left((M-1):M\right)}(s)$ are defined earlier. Thus repeatedly applying this

\begin{align*}
\mathcal{M}_{\hat{\theta}_M}(s) & = 
\exp{\left\{\theta_{0} r^{(1:M)}(s)\right\} }C_M(s) \prod_{m=1}^{M-1}  C^{(m:M)}(s)
\end{align*}
where
\begin{align*}
r^{(m:M)}(s)= \frac{s e^{-(M-m+1)h} 
}{
\left[1 - s(1-e^{-h})   \sum_{l=1}^{\binom{M-m+1}{M-m}} \frac{e^{-(l-1)h}}{\hat{b}_{M-(l-1)}} \right] },
\end{align*}

\begin{align}\label{eq:theo_2_second}
C^{(m:M)}(s)= \left( \frac{ 1  - s \left(1-e^{-h} \right)
 \sum_{l=1}^{\binom{M-m+1}{M-m}} \frac{e^{-(l-1)h}}{\hat{b}_{M-(l-1)}}  
}{ 1  - s \left(1-e^{-h} \right) 
 \sum_{l=1}^{\binom{M-m+1}{M-m}-1} \frac{e^{-(l-1)h}}{\hat{b}_{M-(l-1)}}  
} \right)^{-\hat{a}_m}.
\end{align}
\text{ } 
\end{proof}

\begin{proof}[Proof of \Autoref{theo:coro_mgf}]
From \autoref{theo:theorem_mgf}, define the cumulant-generating function of $\hat{\theta}_M$
\begin{align}\label{eq:cgf}
\mathcal{K}_{\hat{\theta}_M}(s)   & = \log \mathcal{M}_{\hat{\theta}_M}(s) = \theta_{0} r^{(1:M)}(s) + \log C_M(s) + \sum_{m=1}^{M-1} \log C^{(m:M)}(s)
\end{align}
Denote \eqref{eq:theo_2_second} as 
$$C^{(m:M)}(s)= \left( \frac{e_0^{(m:M)}(s)}{e_1^{(m:M)}(s)} \right)^{-\hat{a}_m}.$$
Differentiating $\mathcal{K}_{\hat{\theta}_M}(s)$ in \eqref{eq:cgf}, we find that:
\begin{align}\label{eq:cgf_prime}
\mathcal{K}^\prime_{\hat{\theta}_M}(s) =  \theta_{0} \frac{\partial}{\partial s} r^{(1:M)}(s) + \frac{\partial}{\partial s} \log C_M(s) + \sum_{m=1}^{M-1} \left[ - \hat{a}_m \left( \frac{\partial}{\partial s} \log e_0^{(m:M)}(s) - \frac{\partial}{\partial s} \log e_1^{(m:M)}(s) \right) \right]
\end{align}
where
\begin{align*}
  \frac{\partial}{\partial s} r^{(1:M)}(s)= \frac{e^{-Mh} 
}{
\left[1 - s(1-e^{-h})   \sum_{l=1}^{M} \frac{e^{-(l-1)h}}{\hat{b}_{M-(l-1)}} \right]^2
},
\end{align*}

\begin{align*}
\frac{\partial}{\partial s} \log C_M(s) = \frac{\hat{a}_M ( 1- e^{- h } ) }{\hat{b}_M - s ( 1- e^{- h } ) },
\end{align*}

\begin{align*}
\frac{\partial}{\partial s}  \log e_0^{(m:M)}(s) = \frac{
- (1-e^{-h})   \sum_{l=1}^{\binom{M-m+1}{M-m}} \frac{e^{-(l-1)h}}{\hat{b}_{M-(l-1)}}  
}{
\left[1 - s(1-e^{-h})   \sum_{l=1}^{\binom{M-m+1}{M-m}} \frac{e^{-(l-1)h}}{\hat{b}_{M-(l-1)}} \right]
},
\end{align*}

\begin{align*}
\frac{\partial}{\partial s}  \log e_1^{(m:M)}(s) = \frac{
- (1-e^{-h})   \sum_{l=1}^{\binom{M-m+1}{M-m}-1} \frac{e^{-(l-1)h}}{\hat{b}_{M-(l-1)}} 
}{
\left[1 - s(1-e^{-h})   \sum_{l=1}^{\binom{M-m+1}{M-m}-1} \frac{e^{-(l-1)h}}{\hat{b}_{M-(l-1)}} \right]
}.
\end{align*}

Let $\mathcal{B}_M$ denote the minibatch noise up to iteration $M$. Now taking expectations with respect to the minibatch noise, 
it follows that 
\begin{align*}
    \mathbb{E}\left[\hat{\theta}_M  \right] & = \mathbb{E}\left[\mathbb{E}\left[ \hat{\theta}_M   \mid \mathcal{B}_M \right] \right] =  \mathbb{E}\left[\mathcal{K}^\prime_{\hat{\theta}_M}(0)  \right] \notag \\
    & = \theta_{0} \mathbb{E}\left[ \frac{\partial}{\partial s} r^{(1:M)}(0) \right] + \mathbb{E}\left[ \frac{\partial}{\partial s} \log C_M(0) \right] + \sum_{m=1}^{M-1} \mathbb{E}\left[ \frac{\partial}{\partial s} \log C^{(m:M)}(0) \right] \notag \\
    & = \theta_0  e^{-Mh} + (1-e^{-h})
    \left(a-1 \right) \mathbb{E}\left[ \frac{\hat{a}_M}{\hat{a}_M-1} 
    \right] + (1-e^{-h}) \left(a-1 \right) \sum_{m=1}^{M-1}e^{-(M-m)h} \mathbb{E}\left[ \frac{\hat{a}_m}{\hat{a}_m-1} 
    \right]  \notag \\
    & =  \theta_0 e^{-Mh} + (1-e^{-Mh} ) \left(a-1 \right)  \mathbb{E}\left[\frac{\hat{a}}{\hat{a}-1}   \right]
\end{align*}
where $\mathbb{E}\left[\frac{\hat{a}}{\hat{a}-1} \right]=\mathbb{E}\left[\frac{\hat{a}_m}{\hat{a}_m-1} \right]$ $\forall m=1, \ldots, M$.\\

Now we compute the variance of $\hat{\theta}_M$.
Differentiating $\mathcal{K}^\prime_{\hat{\theta}_M}(s)$ in \eqref{eq:cgf_prime}, we find that:
\begin{align*}
\mathcal{K}^{\prime\prime}_{\hat{\theta}_M}(s) =  \theta_{0} \frac{\partial^2}{\partial s^2} r^{(1:M)}(s) + \frac{\partial^2}{\partial s^2} \log C_M(s) + \sum_{m=1}^{M-1} \left[ - \hat{a}_m \left( \frac{\partial^2}{\partial s^2} \log e_0^{(m:M)}(s) - \frac{\partial^2}{\partial s^2} \log e_1^{(m:M)}(s) \right) \right]
\end{align*}
where

\begin{align*}
    \frac{\partial^2}{\partial s^2} r^{(1:M)}(s)= 
 \frac{ 2e^{-Mh} (1-e^{-h}) \sum_{l=1}^{M} \frac{e^{-(l-1)h}}{\hat{b}_{M-(l-1)}}   }{
  \left[1 - s(1-e^{-h})   \sum_{l=1}^{M} \frac{e^{-(l-1)h}}{\hat{b}_{M-(l-1)}} \right]^3
 },
\end{align*}
\begin{align*}
\frac{\partial^2}{\partial s^2} \log C_M(s) = \frac{\hat{a}_M ( 1- e^{- h } )^2 }{\left[\hat{b}_M - s ( 1- e^{- h } ) \right]^2 },
\end{align*}

\begin{align*}
\frac{\partial^2}{\partial s^2}  \log e_0^{(m:M)}(s) = -\frac{ \left[
 (1-e^{-h})   \sum_{l=1}^{\binom{M-m+1}{M-m}} \frac{e^{-(l-1)h}}{\hat{b}_{M-(l-1)}} \right]^2  
}{
\left[1 - s(1-e^{-h})   \sum_{l=1}^{\binom{M-m+1}{M-m}} \frac{e^{-(l-1)h}}{\hat{b}_{M-(l-1)}} \right]^2
},
\end{align*}

\begin{align*}
\frac{\partial^2}{\partial s^2}  \log e_1^{(m:M)}(s) = -\frac{ \left[
 (1-e^{-h})   \sum_{l=1}^{\binom{M-m+1}{M-m}-1} \frac{e^{-(l-1)h}}{\hat{b}_{M-(l-1)}} \right]^2  
}{
\left[1 - s(1-e^{-h})   \sum_{l=1}^{\binom{M-m+1}{M-m}-1} \frac{e^{-(l-1)h}}{\hat{b}_{M-(l-1)}} \right]^2
}.
\end{align*}

Again taking expectations with respect to the minibatch noise, noting the independence of $\hat{a}_i$ and $\hat{a}_j$ for $i \neq j$, 
it follows that

\begin{align*}
\mathbb{V}\text{ar}\left[\hat{\theta}_M  \right] & = \mathbb{E}\left[\mathbb{V}\text{ar}\left[ \hat{\theta}_M   \mid \mathcal{B}_M \right] \right] =  \mathbb{E}\left[\mathcal{K}^{\prime\prime}_{\hat{\theta}_M}(0)  \right] \notag \\
    & = \theta_{0} \mathbb{E}\left[ \frac{\partial^2}{\partial^2 s} r^{(1:M)}(0) \right] + \mathbb{E}\left[ \frac{\partial^2}{\partial s^2} \log C_M(0) \right] + \sum_{m=1}^{M-1} \mathbb{E}\left[ \frac{\partial^2}{\partial s^2} \log C^{(m:M)}(0) \right] \notag \\
    & = 2 \theta_0 e^{-Mh} (1-e^{-h}) \sum_{l=1}^{M} \mathbb{E}\left[ \frac{e^{-(l-1)h}}{\hat{b}_{M-(l-1)}} \right] + ( 1- e^{- h } )^2 \mathbb{E}\left[ \frac{\hat{a}_M  }{\hat{b}_M^2 } \right] + \notag \\
    & \quad ( 1- e^{- h } )^2 \sum_{m=1}^{M-1} \mathbb{E}\left[ \frac{\hat{a}_m  }{\hat{b}_m^2 } \right]  e^{-2(M-m)h} + 2 ( 1- e^{- h } )^2  \sum_{m=1}^{M-1}  e^{-(M-m)h} \mathbb{E}\left[ \frac{\hat{a}_m  }{\hat{b}_m} \sum_{l=1}^{M-m} \frac{e^{-(l-1)h}}{\hat{b}_{M-(l-1)}} \right]  \notag \\ & =   2 \theta_0 \left( e^{-Mh} - e^{-2Mh} \right) \left( a-1\right)  \mathbb{E}\left[ \frac{1  }{\hat{a}-1 } \right]   + ( 1- e^{- h }  )^2 \left( a-1\right)^2 \mathbb{E}\left[ \frac{\hat{a}  }{\left(\hat{a}-1\right)^2 } \right] \left( \frac{e^{2h}-e^{-2(M-1)h}}{e^{2h}-1} \right) + \notag \\ &
    \quad 2 ( 1- e^{- h }  )^2 \left( a-1\right)^2 \mathbb{E}\left[ \frac{\hat{a}  }{\hat{a}-1 } \right] \mathbb{E}\left[ \frac{1  }{\hat{a}-1 } \right] \left[ \frac{e^{-(M-1)h}-1 }{\left( 1-e^{-h}\right) \left( 1-e^{h}\right) } - \frac{e^{-2(M-1)h}-1}{\left( 1-e^{-h}\right) \left( 1-e^{2h}\right)} \right] \\
    & =  2 \theta_0 \left( e^{-Mh} - e^{-2Mh} \right) \left( a-1\right)  \mathbb{E}\left[ \frac{1  }{\hat{a}-1 } \right]   + ( 1- e^{- h }  )^2 \left( a-1\right)^2 \mathbb{E}\left[ \frac{\hat{a}  }{\left(\hat{a}-1\right)^2 } \right] 
      \left( \frac{1-e^{-2Mh}}{1-e^{-2h}} \right) + 
    \notag \\ &
    \quad 2 ( 1- e^{- h }  )^2 \left( a-1\right)^2 \mathbb{E}\left[ \frac{\hat{a}  }{\hat{a}-1 } \right] \mathbb{E}\left[ \frac{1  }{\hat{a}-1 } \right] \left[ \frac{e^{-Mh}-e^{-h} }{\left( 1-e^{-h}\right) \left( e^{-h}-1\right) } - \frac{e^{-2Mh}-e^{-2h}}{\left( 1-e^{-h}\right) \left( e^{-2h}-1\right)} \right]
\end{align*}    
where $\mathbb{E}\left[\frac{1}{\hat{a}-1} \right]=\mathbb{E}\left[\frac{1}{\hat{a}_m-1} \right]$ and $\mathbb{E}\left[ \frac{\hat{a}  }{\left(\hat{a}-1\right)^2 } \right]=\mathbb{E}\left[ \frac{\hat{a}_m  }{\left(\hat{a}_m-1\right)^2 } \right]$ $\forall m=1, \ldots, M$.
\end{proof}

\section{Appendix 2}\label{app:app3}
Here we provide an alternative parametrization of the control variate-based stochastic Cox-Ingersoll-Ross process that still approximatively targets the desired posterior. 

\subsection*{Alternative parametrization}
A parametrization equivalent to \ref{eq:scir_cv} that still approximatively targets the desired posterior distribution $\operatorname{Gamma}\left(a, 1\right)$ is achieved by setting $a=\frac{\hat{a}}{\hat{b}}$, $b=\hat{b}$, $\sigma^2=2$ in \eqref{eq:st_cir}. Thus, the resulting process in differential form is:
\begin{align}\label{eq:scir_cv_rep}
\mathrm{d} \theta_t=\hat{b}\left(\frac{\hat{a}}{\hat{b}}-\theta_t\right) \mathrm{d} t+ \sqrt{2\theta_t} \mathrm{d} W_t.
\end{align}
The diffusion in \autoref{eq:scir_cv_rep} has the following transition distribution
\begin{align}\label{eq:scir_cv_trans_rep}
\hat{\theta}_{m+1} \mid \hat{\theta}_m=\vartheta_m \sim \frac{1-e^{-\hat{b}_m h}}{2\hat{b}_m} W, \quad W \sim \chi^2\left(2 \hat{a}_m, 2 \vartheta_m \frac{\hat{b}_m e^{-\hat{b}_m h}}{1-e^{-\hat{b}_m h}}\right)
\end{align}
where $\hat{a}=\alpha + N / n \sum_{i \in \mathcal{S}} z_i$ and $\hat{b}=\frac{\hat{a}-1}{a-1}$. As mentioned in \Autoref{sec:section4}, \eqref{eq:scir_cv_rep} allows for values of $\hat{b}<0$. Nevertheless, to apply the control variate framework, $a$ needs to be known, as it appears in $\hat{b}$. In practice, when the $z$ are latent variables, these can be estimated every, say, $\ell$ iterations.\\

In the following, we derive the moment-generating function of $\hat{\theta}_m$ under this parametrization. The result is formally stated in \autoref{theo:theorem_mgf_2}.

\begin{theorem}\label{theo:theorem_mgf_2}
Let $\hat{\theta}_M$ be generated by the control variate-based stochastic Cox--Ingersoll--Ross process defined in \eqref{eq:scir_cv_trans_rep} starting from $\theta_0$ after $M$ steps with stepsize $h$. Then the moment-generating function of $\hat{\theta}_M$ is 
\begin{align*}
\mathcal{M}_{\hat{\theta}_M}(s) & = MGF_{\hat{\theta}_M \mid \theta_{0}}(s) =  \exp{\left\{\theta_{0} r^{(1:M)}(s)\right\} }C_M(s) \prod_{m=1}^{M-1}  C^{(m:M)}(s)
\end{align*}
where
\begin{align*}
r^{(m:M)}(s)= \frac{s e^{-h\sum_{j=m}^M \hat{b}_j }  
}{
1 - s \left[ e^{-h\sum_{j=m}^M \hat{b}_j }  
\sum_{l=1}^{\binom{M-m+1}{M-m}} 
\frac{(1-e^{-h \hat{b}_{M-(l-1)} })}{\hat{b}_{M-(l-1)}} e^{h\sum_{j=m}^{M-(l-1)} \hat{b}_j}
\right] },
\end{align*}
\begin{align*}
C_m(s)=\left[ \frac{ \hat{b}_m-s(1-e^{-h\hat{b}_m })  }{\hat{b}_m} \right]^{-\hat{a}_m}, 
\end{align*}
\begin{align}\label{eq:theo_2_first}
C^{(m:M)}(s)= \left( \frac{ 
1 - s \left[ 
e^{-h\sum_{j=m}^M \hat{b}_j }
\sum_{l=1}^{\binom{M-m+1}{M-m}} 
\frac{(1-e^{-h \hat{b}_{M-(l-1)} })}{\hat{b}_{M-(l-1)}} e^{h\sum_{j=m}^{M-(l-1)} \hat{b}_j} 
\right] }
{ 
1 - s \left[ 
e^{-h\sum_{j=m}^M \hat{b}_j } 
\sum_{l=1}^{\binom{M-m+1}{M-m}-1} 
\frac{(1-e^{-h \hat{b}_{M-(l-1)} })}{\hat{b}_{M-(l-1)}} e^{h\sum_{j=m}^{M-(l-1)} \hat{b}_j} 
\right] 
} \right)^{-\hat{a}_m}.
\end{align}
\end{theorem}

Deriving the moment-generating function enables us to find the non-asymptotic bias and variance of the control variate-based stochastic Cox-Ingersoll-Ross process, which are more interpretable than
the moment-generating function itself. The results are stated formally in the following \Autoref{theo:coro_mgf_2}.
\begin{corollary}\label{theo:coro_mgf_2}
Given the setup of \autoref{theo:theorem_mgf_2}, it follows that
\begin{align}\label{eq:mean_scir_cv_2}
        \mathbb{E}\left[\hat{\theta}_M  \right] 
 & =  \theta_0 e^{-Mt} \left[ 
\mathcal{M}_{\hat{a}}\left( t \right)
 \right]^M + \frac{1-e^{-Mt} \left[ 
\mathcal{M}_{\hat{a}}\left( t \right)
 \right]^M 
 }{1-e^{-t}  
\mathcal{M}_{\hat{a}}\left( t \right)
 }\mathbb{E}\left[ \hat{a}\frac{(1-e^{-h\hat{b}} )}{\hat{b}}     \right], 
\end{align}
\begin{align}\label{eq:variance_scir_cv_2}
    \mathbb{V}ar\left[\hat{\theta}_M  
    \right]
    & = 
2 \mathbb{E}\left[ \frac{(e^{-h\hat{b} }-e^{-2h\hat{b} })}{\hat{b} }  \right] \left\{ \theta_0 \mathcal{C}_1 + \mathbb{E}\left[ \hat{a}  \frac{ ( 1-e^{-h\hat{b}  } )  }{\hat{b} } \right] \mathcal{C}_3 
\right\} + \mathbb{E}\left[ \hat{a}
    \left( \frac{  1- e^{-\hat{b} h }   }{\hat{b} } \right)^2  \right] \mathcal{C}_2,
\end{align}
where $\mathcal{M}_{\hat{a}}\left( t \right)= \mathbb{E}\left[ e^{t\hat{a} }\right]$ is the moment-generating function of $\hat{a}$, $t=-\frac{h}{a-1}$ and 
\begin{align*}
    & \mathcal{C}_1= \frac{ 
       e^{-Mt} \left[ 
       \mathcal{M}_{\hat{a}}\left( t\right)
       \right]^{M}- e^{-2Mt} \left[ 
       \mathcal{M}_{\hat{a}}\left( 2t \right)
       \right]^{M} 
       }{
       e^{-t}  
       \mathcal{M}_{\hat{a}}\left( t \right) - e^{-2t}  \mathcal{M}_{\hat{a}}\left( 2t \right)}, \quad
\mathcal{C}_2=\frac{1-e^{-2Mt}\left[
    \mathcal{M}_{\hat{a}}\left( 2t \right)
    \right]^M}{1-e^{-2t}
    \mathcal{M}_{\hat{a}}\left( 2t \right) }, \\
    & \mathcal{C}_3=    \left\{   
     1 +
     e^{-t} \mathcal{M}_{\hat{a}}\left( t\right) +  e^{-2t} \mathcal{M}_{\hat{a}}\left( 2t \right) +  
    \frac{ e^{-Mt}\left[ \mathcal{M}_{\hat{a}}\left( t \right) \right]^{M}-  e^{-3t} \left[ \mathcal{M}_{\hat{a}}\left( t \right) \right]^{3}}{ \left[
     e^{-t}
     \mathcal{M}_{\hat{a}}\left( t \right)
     - e^{-2t}
    \mathcal{M}_{\hat{a}}\left( 2t \right) 
    \right] \left[  e^{-t} \mathcal{M}_{\hat{a}}\left( t \right) -1 \right]} -
          \right. 
    \notag \\ &
    \left. \qquad \quad   
    \frac{  e^{-2Mt} \left[\mathcal{M}_{\hat{a}}\left( 2t \right) \right]^{M}-  e^{-6t} \left[ \mathcal{M}_{\hat{a}}\left( 2t \right) \right]^{3}}{ \left[ e^{-t} \mathcal{M}_{\hat{a}}\left( t\right)- e^{-2t} \mathcal{M}_{\hat{a}}\left( 2t \right) \right] \left[e^{-2t}\mathcal{M}_{\hat{a}}\left( 2t \right) -1\right]} 
    \right\}.  
\end{align*}
\end{corollary}
Thus, the expectation $\mathbb{E}\left[\hat{\theta}_M  \right] $ and variance $\mathbb{V}ar\left[\hat{\theta}_M  
\right]$ depend on the distribution
of the random generating mechanism of the mini-batch estimator $\hat{a}$ through the expectation of non-linear functions we define as $\phi\left( \cdot \right)$, including its moment-generating function $\mathcal{M}_{\hat{a}}\left( t \right)$.

Since $\mathbb{E}_\pi\left[\theta\right]=a$, expanding expectations of the type $\mathbb{E}\left[\phi\left( \hat{a} \right)\right]$ at $\mathbb{E}\left[ \hat{a}\right]=a$ and assuming the first two moments of $\hat{a}$ properly summarize its distributional properties, it follows
\begin{align*}
    \lim_{M \rightarrow \infty}  \Big| \frac{1}{M}\sum_{m=1}^M \mathbb{E}\left[\hat{\theta}_m\right]- \mathbb{E}_\pi\left[\theta\right] \Big| \approx \frac{1-e^{-h}(1+h)}{1-e^{-h}(1+0.5h^2 \sigma^2(a) )} \sigma^2(a)
\end{align*}
where $\sigma^2(a)= (a-1)^{-2} \mathbb{V}\text{ar}\left[\hat{a}  \right]$. The asymptotic behavior of the variance can be analyzed in a similar way and will again depend on $\sigma^2(a)$.

This result shows that, while the accuracy of the process can be improved for $M$ sufficiently large, there is a systematic component due to the stochastic noise intrinsic in the estimate $\hat{a}$ and the overall asymptotic accuracy is inflated relative to the exact underlying Cox-Ingersoll-Ross process. 
Nevertheless, the extra term is substantially lower than that of the stochastic Cox-Ingersoll-Ross process of \cite{Baker2018LargeScaleSS}, as we investigate in the next, and it further benefits from non-sparse settings, as evident from the expression of $\sigma^2(a)$ and seen in \autoref{fig:run_ex_1}.

\begin{remark}\label{rem1}
Let $\mathcal{S}$ be a simple random sample of $\{1, \ldots, N\}$ without replacement and with $|\mathcal{S}|=n$, then 
\begin{align}\label{eq:hyp_geo}
    \overline{Z}=\sum_{i \in \mathcal{S}} z_i \sim \operatorname{HyperGeo}\left(N, Np, n \right), \quad p=\frac{1}{N}\sum_{i=1}^N z_i
\end{align}
It follows: 
\begin{align}\label{eq:var_scir_3}
\mathbb{V}\text{ar}\left[\hat{a}  \right] = \frac{N^2}{n^2} \mathbb{V}\text{ar}\left[ \overline{Z}  \right] = \frac{N^2}{n}p \left( 1-p\right) \frac{N-n}{N-1}
\end{align}
and
\begin{align}\label{eq:mgf_a_hat_simple}
   \mathcal{M}_{\hat{a}}\left( t \right)= e^{\alpha t} \mathcal{M}_{\overline{Z}}\left( tN/n  \right), 
\end{align}
where $M_{\overline{Z}}\left(s \right)$ is the moment generating function of $\overline{Z}$, given by
\begin{align}\label{eq:mgf_Z}
    M_{\overline{Z}}\left(s \right) =   \mathbb{E}\left[ e^{ s \overline{Z} } 
\right] =\frac{\binom{N-Np}{n} \hypgeo{2}{1}
\left(-n, -Np, N-Np-n+1; e^{s} \right)  }{\binom{N}{n}}
\end{align}
and $\hypgeo{2}{1}
\left(a, b, c; z \right)$ is the ordinary hypergeometric function.
\end{remark}

It is instructive to outline the gain achieved by the control variate-based stochastic Cox-Ingersoll-Ross process over the stochastic Cox-Ingersoll-Ross process of \cite{Baker2018LargeScaleSS} in terms of variance reduction.\\
\cite{Baker2018LargeScaleSS} showed that for the stochastic Cox-Ingersoll-Ross process it holds:
\begin{align}\label{eq:var_scir_1}
    \mathbb{V}\text{ar}\left[\hat{\theta}_M  \right] = \mathbb{V}\text{ar}\left[\theta_M  \right] + (1-e^{-2Mh} ) \frac{1-e^{-h}}{1+e^{-h}}  \mathbb{V}\text{ar}\left[\hat{a}  \right]
\end{align}

\begin{figure}[!h]
\hspace{-3.5em}
\subfloat[Variance of the \emph{main} parametrization as in \eqref{eq:variance}.]{
\includegraphics[clip,trim={5cm 0cm 5cm 0cm},width=0.6\textwidth]
{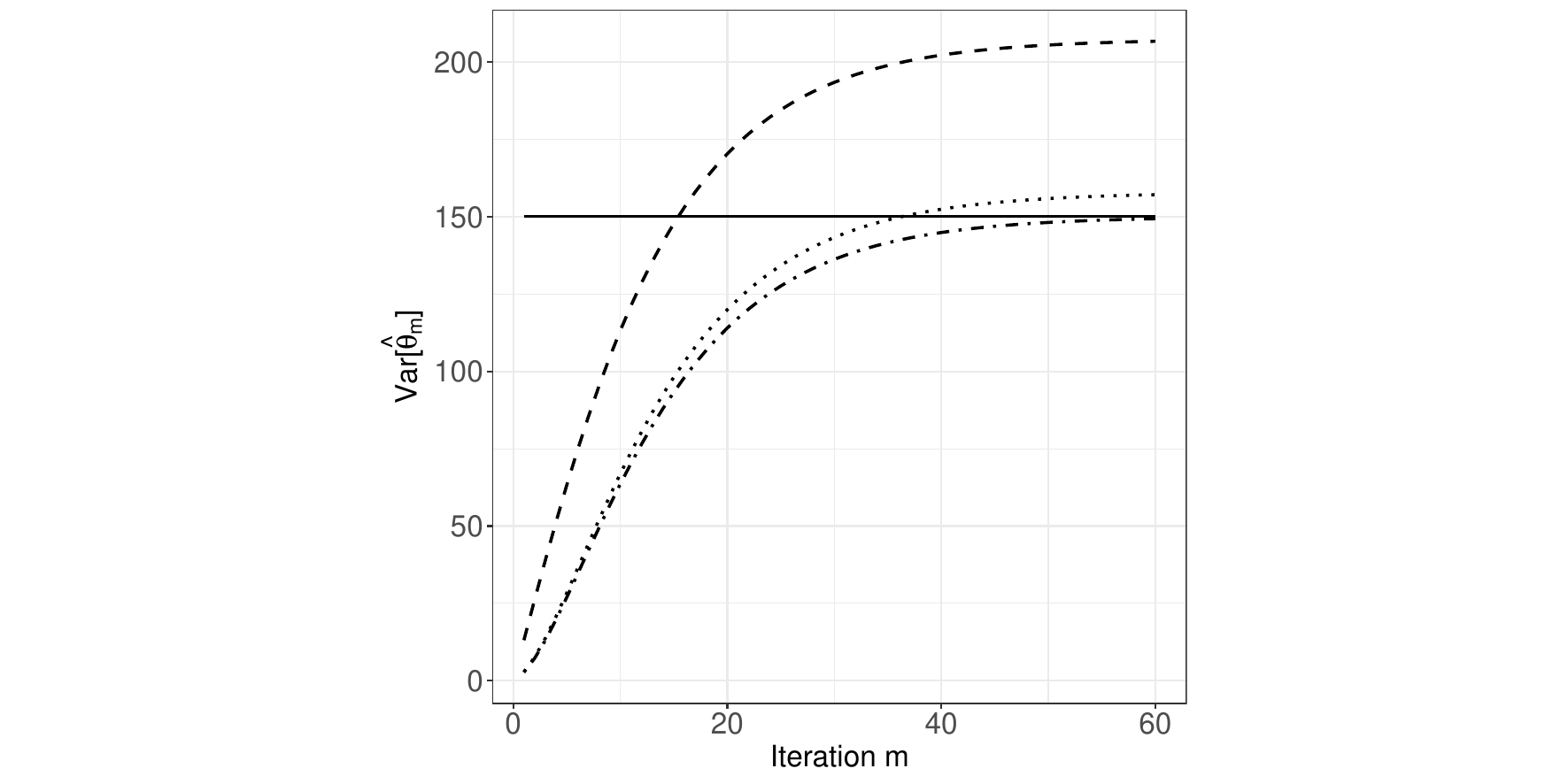}} \hspace{-3em}
\subfloat[Variance of the alternative parametrization as in \eqref{eq:variance_scir_cv_2}.]{
\includegraphics[clip,trim={5cm 0cm 5cm 0cm},width = 0.6\textwidth]
{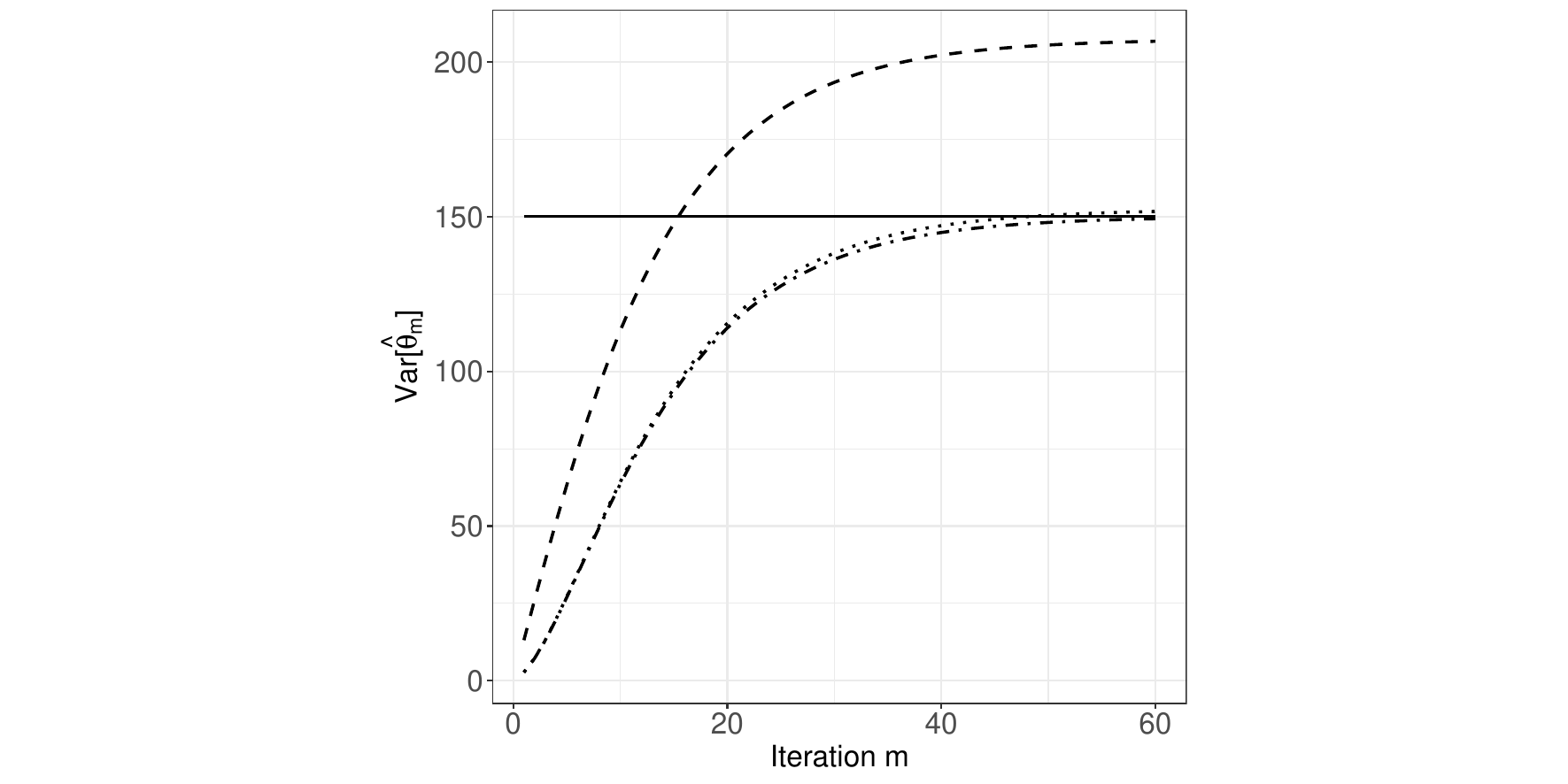}}
\caption{Variance of: the stochastic Cox-Ingersoll-Ross process of \cite{Baker2018LargeScaleSS} as in \eqref{eq:var_scir_1}
(dashed line); the control variate-based stochastic Cox-Ingersoll-Ross process 
(dotted line);
the exact underlying Cox-Ingersoll-Ross process as in \eqref{eq:exact_cir_moments} (dot-dash line); the posterior variance $\mathbb{V}\text{ar}_{\pi}\left[ \theta  \right]=a$ (solid line). }\label{fig:var_comparison}
\end{figure}

\autoref{fig:var_comparison} illustrates the advantage of using the control variate version of the stochastic Cox-Ingersoll-Ross process. While the variance of the stochastic Cox-Ingersoll-Ross process is somewhat inflated over the variance of the exact underlying Cox-Ingersoll-Ross process, the control variate-based version reduces the extra term involving $\mathbb{V}\text{ar}\left[\hat{a}  \right]$ getting substantially closer to $\mathbb{V}\text{ar}\left[\theta_M  \right]$. 
Moreover, it can be appreciated how the alternative parametrization is more accurate and empirically displays an almost identical level of accuracy as a full-data exact Cox--Ingersoll--Ross process. We believe this can be explained because the alternative parametrization involves 
the moment-generating function of the mini-batch estimator $\hat{a}$ which embeds distributional properties that are not captured by simple expectations of non-linear functions of $\hat{a}$. 
For this illustration, we set: $\theta_0=7.67$,  $\alpha=0.1$, $h=0.1$, $N=1000$, $p=0.15$, $n=100$. While for the variance of the stochastic Cox-Ingersoll-Ross process, $\mathbb{V}\text{ar}\left[\hat{a} \right]$ is specialized with \eqref{eq:var_scir_3} and, for the variance of the alternative parametrization of the control variate-based stochastic Cox-Ingersoll-Ross process, $\mathcal{M}_{\hat{a}}\left( t \right)$ is specialized with  \eqref{eq:mgf_a_hat_simple}--\eqref{eq:mgf_Z}, expectations of the type $\mathbb{E}\left[ \phi\left( \hat{a} \right) \right]$ for other non-linear functions $\phi\left( \cdot \right)$ as in \eqref{eq:variance} and \eqref{eq:variance_scir_cv_2} are evaluated via Monte Carlo method with $1\, 000$ samples under \eqref{eq:hyp_geo}.

We now provide proofs of \autoref{theo:theorem_mgf_2} and \Autoref{theo:coro_mgf_2}.

\begin{proof}[Proof of \autoref{theo:theorem_mgf_2}]
The proof follows to recursively applying the same 
properties of the non-central chi-squared distribution
as in the Proof of \autoref{theo:theorem_mgf} in \Autoref{app:app1}, after re-defining the following quantities
\begin{align}\label{eq:supp_eq_2}
r_m(s)= \frac{s \hat{b}_m e^{-\hat{b}_m h}}{\hat{b}_m-s(1-e^{-\hat{b}_m h})}, \qquad C_m(s)=\left( \frac{ \hat{b}_m-s(1-e^{-\hat{b}_m h})  }{\hat{b}_m} \right)^{-\hat{a}_m} 
\end{align}
for any step $m=1, 2, \ldots, M$.
Note that, contrary to \ref{eq:supp_eq}, $\hat{b}$ now also appears in $e^{-h}$ in \eqref{eq:supp_eq_2}.
\end{proof}

\begin{proof}[Proof of \Autoref{theo:coro_mgf_2}]
From \autoref{theo:theorem_mgf_2}, define the cumulant-generating function of $\hat{\theta}_M$
\begin{align}\label{eq:cgf_2}
\mathcal{K}_{\hat{\theta}_M}(s)   & = \log \mathcal{M}_{\hat{\theta}_M}(s) = \theta_{0} r^{(1:M)}(s) + \log C_M(s) + \sum_{m=1}^{M-1} \log C^{(m:M)}(s)
\end{align}
Denote \eqref{eq:theo_2_first} as 
$$C^{(m:M)}(s)= \left( \frac{e_0^{(m:M)}(s)}{e_1^{(m:M)}(s)} \right)^{-\hat{a}_m}.$$
Differentiating $\mathcal{K}_{\hat{\theta}_M}(s)$ in \eqref{eq:cgf_2}, we find that:
\begin{align}\label{eq:cgf_prime_2}
\mathcal{K}^\prime_{\hat{\theta}_M}(s) =  \theta_{0} \frac{\partial}{\partial s} r^{(1:M)}(s) + \frac{\partial}{\partial s} \log C_M(s) + \sum_{m=1}^{M-1} \left[ - \hat{a}_m \left( \frac{\partial}{\partial s} \log e_0^{(m:M)}(s) - \frac{\partial}{\partial s} \log e_1^{(m:M)}(s) \right) \right]
\end{align}
where
\begin{align*}
    \frac{\partial}{\partial s} r^{(1:M)}(s) =  
 \frac{ 
 e^{-h\sum_{j=1}^M \hat{b}_j }  }{
\left\{
1 - s \left[ e^{-h\sum_{j=1}^M \hat{b}_j }  
\sum_{l=1}^{M} 
\frac{(1-e^{-h \hat{b}_{M-(l-1)} })}{\hat{b}_{M-(l-1)}} e^{h\sum_{j=1}^{M-(l-1)} \hat{b}_j}
\right]
\right\}^2
 },
\end{align*}

\begin{align*}
\frac{\partial}{\partial s} \log C_M(s) = \frac{\hat{a}_M ( 1- e^{-\hat{b}_M h } ) }{\hat{b}_M - s ( 1- e^{-\hat{b}_M h } ) },
\end{align*}

\begin{align*}
\frac{\partial}{\partial s}  \log e_0^{(m:M)}(s) = \frac{
- e^{-h\sum_{j=m}^M \hat{b}_j }    
\sum_{l=1}^{\binom{M-m+1}{M-m}} 
\frac{(1-e^{-h \hat{b}_{M-(l-1)} })}{\hat{b}_{M-(l-1)}} e^{h\sum_{j=m}^{M-(l-1)} \hat{b}_j} 
}{  
1 - s \left[ 
e^{-h\sum_{j=m}^M \hat{b}_j }  
\sum_{l=1}^{\binom{M-m+1}{M-m}} 
\frac{(1-e^{-h \hat{b}_{M-(l-1)} })}{\hat{b}_{M-(l-1)}} e^{h\sum_{j=m}^{M-(l-1)} \hat{b}_j} 
\right]
},
\end{align*}

\begin{align*}
\frac{\partial}{\partial s}  \log e_1^{(m:M)}(s) = \frac{
- e^{-h\sum_{j=m}^M \hat{b}_j }  
\sum_{l=1}^{\binom{M-m+1}{M-m}-1} 
\frac{(1-e^{-h \hat{b}_{M-(l-1)} })}{\hat{b}_{M-(l-1)}} e^{h\sum_{j=m}^{M-(l-1)} \hat{b}_j} 
}{ 
1 - s \left[ 
e^{-h\sum_{j=m}^M \hat{b}_j }   
\sum_{l=1}^{\binom{M-m+1}{M-m}-1} 
\frac{(1-e^{-h \hat{b}_{M-(l-1)} })}{\hat{b}_{M-(l-1)}} e^{h\sum_{j=m}^{M-(l-1)} \hat{b}_j} 
\right]
}.
\end{align*}
Let $\mathcal{B}_M$ denote the minibatch noise up to iteration $M$. Now taking expectations with respect to the minibatch noise, noting the independence of $\hat{a}_i$ and $\hat{a}_j$ for $i \neq j$, 
it follows that 
\begin{align*}
    \mathbb{E}\left[\hat{\theta}_M  \right] & = \mathbb{E}\left[\mathbb{E}\left[ \hat{\theta}_M   \mid \mathcal{B}_M \right] \right] =  \mathbb{E}\left[\mathcal{K}^\prime_{\hat{\theta}_M}(0)  \right] \notag \\
    & = \theta_{0} \mathbb{E}\left[ \frac{\partial}{\partial s} r^{(1:M)}(0) \right] + \mathbb{E}\left[ \frac{\partial}{\partial s} \log C_M(0) \right] + \sum_{m=1}^{M-1} \mathbb{E}\left[ \frac{\partial}{\partial s} \log C^{(m:M)}(0) \right] \notag \\
& = \theta_0  \mathbb{E}\left[ e^{-h\sum_{m=1}^M \hat{b}_m } 
    \right] + 
   \mathbb{E}\left[ \frac{\hat{a}_M}{\hat{b}_M} \left(1-e^{-h\hat{b}_M} \right)
    \right] + 
\sum_{m=1}^{M-1} \left\{ \mathbb{E}\left[ \frac{\hat{a}_m}{\hat{b}_m} \left(1-e^{-h\hat{b}_m} \right) 
\right] \times \mathbb{E}\left[ e^{-h\sum_{j=m+1}^M \hat{b}_j} 
\right]  \right\} \notag \\
& = \theta_0 \prod_{m=1}^M  \mathbb{E}\left[ e^{-h\hat{b}_m }\right] + \mathbb{E}\left[ \frac{\hat{a}_M}{\hat{b}_M} \left(1-e^{-h\hat{b}_M} \right)    \right] +  \sum_{m=1}^{M-1} \left\{ \mathbb{E}\left[ \frac{\hat{a}_m}{\hat{b}_m} \left(1-e^{-h\hat{b}_m} \right) 
\right] \times \prod_{j=m+1}^M \mathbb{E}\left[ e^{-h\hat{b}_j} 
\right]  \right\} \notag \\
& =  \theta_0 \left( \mathbb{E}\left[ e^{-h\hat{b} }\right] \right)^M +  \left\{ 1 + \sum_{m=1}^{M-1}  \left( \mathbb{E}\left[ e^{-h\hat{b} }\right] \right)^{M-m}   \right\} \mathbb{E}\left[ \frac{\hat{a}}{\hat{b}} \left(1-e^{-h\hat{b}} \right)    \right] \notag \\
& =  \theta_0 \left( \mathbb{E}\left[ e^{-h\hat{b} }\right] \right)^M + \frac{1-\left( \mathbb{E}\left[ e^{-h\hat{b} }\right] \right)^M}{1-\mathbb{E}\left[ e^{-h\hat{b} }\right] }\mathbb{E}\left[ \frac{\hat{a}}{\hat{b}} \left(1-e^{-h\hat{b}} \right)    \right] \notag \\
& =  \theta_0 e^{\frac{Mh}{a-1}} \left[ 
\mathcal{M}_{\hat{a}}\left( -\frac{h}{a-1} \right)
 \right]^M + \frac{1-e^{\frac{Mh}{a-1}} \left[ 
\mathcal{M}_{\hat{a}}\left( -\frac{h}{a-1} \right)
 \right]^M 
 }{1-e^{\frac{h}{a-1}}  
\mathcal{M}_{\hat{a}}\left( -\frac{h}{a-1} \right)
 }\mathbb{E}\left[ \hat{a} \frac{(1-e^{-h\hat{b}}) }{\hat{b}}      \right] 
\end{align*}
where $\mathcal{M}_{\hat{a}}\left( t \right)= \mathbb{E}\left[ e^{t\hat{a} }\right]$ is the moment-generating function of $\hat{a}$.\\

Now we compute the variance of $\hat{\theta}_M$.
Differentiating $\mathcal{K}^\prime_{\hat{\theta}_M}(s)$ in \eqref{eq:cgf_prime_2}, we find that:
\begin{align*}
\mathcal{K}^{\prime\prime}_{\hat{\theta}_M}(s) =  \theta_{0} \frac{\partial^2}{\partial s^2} r^{(1:M)}(s) + \frac{\partial^2}{\partial s^2} \log C_M(s) + \sum_{m=1}^{M-1} \left[ - \hat{a}_m \left( \frac{\partial^2}{\partial s^2} \log e_0^{(m:M)}(s) - \frac{\partial^2}{\partial s^2} \log e_1^{(m:M)}(s) \right) \right]
\end{align*}
where
\begin{align*}
    \frac{\partial^2}{\partial s^2} r^{(1:M)}(s)= 
 \frac{ 2
 e^{-2h\sum_{j=1}^M \hat{b}_j }  \sum_{l=1}^{M} 
 \frac{(1-e^{-h\hat{b}_{M-(l-1)} })}{\hat{b}_{M-(l-1)}} e^{h\sum_{j=1}^{M-(l-1)}} \hat{b}_j   }{
  \left\{
1- s \left[ e^{-h\sum_{j=1}^M \hat{b}_j }  
\sum_{l=1}^{M} 
\frac{(1-e^{-h \hat{b}_{M-(l-1)} })}{\hat{b}_{M-(l-1)}} e^{h\sum_{j=1}^{M-(l-1)} \hat{b}_j}
\right]
\right\}^3
 },
\end{align*}
\begin{align*}
\frac{\partial^2}{\partial s^2} \log C_M(s) = \frac{\hat{a}_M ( 1- e^{-\hat{b}_M h } )^2 }{\left[\hat{b}_M - s ( 1- e^{-\hat{b}_M h } ) \right]^2 },
\end{align*}

\begin{align*}
\frac{\partial^2}{\partial s^2}  \log e_0^{(m:M)}(s) = -\frac{
\left[
e^{-h\sum_{j=m}^M \hat{b}_j }  
\sum_{l=1}^{\binom{M-m+1}{M-m}} 
\frac{(1-e^{-h \hat{b}_{M-(l-1)} })}{\hat{b}_{M-(l-1)}} e^{h\sum_{j=m}^{M-(l-1)} \hat{b}_j} 
\right]^2
}{
\left\{  
1- s \left[ 
e^{-h\sum_{j=m}^M \hat{b}_j }  
\sum_{l=1}^{\binom{M-m+1}{M-m}} 
\frac{(1-e^{-h \hat{b}_{M-(l-1)} })}{\hat{b}_{M-(l-1)}} e^{h\sum_{j=m}^{M-(l-1)} \hat{b}_j} 
\right]
\right\}^2
},
\end{align*}

\begin{align*}
\frac{\partial^2}{\partial s^2}  \log e_1^{(m:M)}(s) =- \frac{ 
\left[
e^{-h\sum_{j=m}^M \hat{b}_j }  
\sum_{l=1}^{\binom{M-m+1}{M-m}-1} 
\frac{(1-e^{-h \hat{b}_{M-(l-1)} })}{\hat{b}_{M-(l-1)}} e^{h\sum_{j=m}^{M-(l-1)} \hat{b}_j} 
\right]^2
}{
\left\{
1- s \left[ 
e^{-h\sum_{j=m}^M \hat{b}_j }  
\sum_{l=1}^{\binom{M-m+1}{M-m}-1} 
\frac{(1-e^{-h \hat{b}_{M-(l-1)} })}{\hat{b}_{M-(l-1)}} e^{h\sum_{j=m}^{M-(l-1)} \hat{b}_j} 
\right]
\right\}^2
}.
\end{align*}

Again taking expectations with respect to the minibatch noise, noting the independence of $\hat{a}_i$ and $\hat{a}_j$ for $i \neq j$, 
it follows that
{\allowdisplaybreaks
\begin{align}
\mathbb{V}\text{ar}\left[\hat{\theta}_M  \right] & = \mathbb{E}\left[\mathbb{V}\text{ar}\left[ \hat{\theta}_M   \mid \mathcal{B}_M \right]  \right] =  \mathbb{E}\left[\mathcal{K}^{\prime\prime}_{\hat{\theta}_M}(0)  \right] \label{eq:var_second_exp}   \\
    & = \theta_{0} \mathbb{E}\left[ \frac{\partial^2}{\partial^2 s} r^{(1:M)}(0) \right] + \mathbb{E}\left[ \frac{\partial^2}{\partial s^2} \log C_M(0) \right] + \sum_{m=1}^{M-1} \mathbb{E}\left[ \frac{\partial^2}{\partial s^2} \log C^{(m:M)}(0) \right] \notag \\
    & = 2\theta_0 
    \underbrace{
    \mathbb{E}\left[
    e^{-2h\sum_{j=1}^M \hat{b}_j }  \sum_{l=1}^{M} \frac{(1-e^{-h\hat{b}_{M-(l-1)}})}{\hat{b}_{M-(l-1)}} e^{h \sum_{j=1}^{M-(l-1)}  \hat{b}_j}  \right]}_{E_1} + \label{eq:var_second_1} \\
    & \quad
    \underbrace{
\mathbb{E}\left[
    \frac{\hat{a}_M  }{\hat{b}_M^2 } ( 1- e^{-\hat{b}_M h } )^2 \right] +   \sum_{m=1}^{M-1} 
    \mathbb{E}\left[
    \frac{\hat{a}_m  }{\hat{b}_m^2 } ( 1- e^{-\hat{b}_m h } )^2   e^{-2h\sum_{j=m+1}^M \hat{b}_j } \right]}_{E_2} + \label{eq:var_second_2}
    \\
    & \quad 2
    \underbrace{
    \sum_{m=1}^{M-1}
    \mathbb{E}\left[  \frac{\hat{a}_m  }{\hat{b}_m } ( 1- e^{-\hat{b}_m h } ) e^{-2h\sum_{j=m+1}^M \hat{b}_j } \left( 
    \sum_{l=1}^{M-m} \frac{(1-e^{-h\hat{b}_{M-(l-1)}})}{\hat{b}_{M-(l-1)}} e^{h \sum_{j=m+1}^{M-(l-1)} \hat{b}_j}
    \right) \right]}_{E_3}  \label{eq:var_second_3}
\end{align}  
}
We work out quantities $E_1$, $E_2$ and $E_3$ in \eqref{eq:var_second_1}--\eqref{eq:var_second_3} separetely in the following.\\

As for $E_1$, first separate out the terms for $l \in \{ 1, M \}$. Then, note that 
$$\left(2\sum_{j=1}^M \hat{b}_j -\sum_{j=1}^{M-(l-1)}  \hat{b}_j\right)=\left(\sum_{j=1}^M \hat{b}_j +\sum_{j=M-(l-2)}^{M}  \hat{b}_j\right)=\left(\sum_{j=1}^{M-l} \hat{b}_j +2\sum_{j=M-(l-2)}^{M}  \hat{b}_j + \hat{b}_{M-(l-1)}\right)$$

{\allowdisplaybreaks
\begin{align}\label{eq:var_E1}
E_1 &=  \mathbb{E}\left[
    e^{-2h\sum_{j=1}^M \hat{b}_j }  \sum_{l=1}^{M} \frac{(1-e^{-h\hat{b}_{M-(l-1)}})}{\hat{b}_{M-(l-1)}} e^{h \sum_{j=1}^{M-(l-1)}  \hat{b}_j}  \right] =\sum_{l=1}^{M} \mathbb{E}\left[ \frac{(1-e^{-h\hat{b}_{M-(l-1)}})}{\hat{b}_{M-(l-1)}}  e^{-h\left(2\sum_{j=1}^M \hat{b}_j -\sum_{j=1}^{M-(l-1)}  \hat{b}_j\right) } 
      \right]  \notag \\
     &= \mathbb{E}\left[ \frac{(1-e^{-h\hat{b}_{M}})}{\hat{b}_{M}}  e^{-h\sum_{j=1}^M \hat{b}_{j} } 
      \right]  + \sum_{l=2}^{M-1} \mathbb{E}\left[ \frac{(1-e^{-h\hat{b}_{M-(l-1)}})}{\hat{b}_{M-(l-1)}}  e^{-h\left(\sum_{j=1}^M \hat{b}_j +\sum_{j=M-(l-2)}^{M}  \hat{b}_j\right) } 
      \right] 
       + \notag \\
       & \quad \;
\mathbb{E}\left[ \frac{(e^{-h\hat{b}_{1}}-e^{-2h\hat{b}_{1}})}{\hat{b}_{1}}  e^{-2h\sum_{j=2}^M \hat{b}_j } 
      \right] \notag \\
      &= \left\{  \mathbb{E}\left[ \frac{(e^{-h\hat{b}_{M}}-e^{-2h\hat{b}_{M}})}{\hat{b}_{M}}  \right]
    \prod_{j=1}^{M-1}
    \mathbb{E}\left[ e^{-h\hat{b}_{j}} 
      \right]  + \sum_{l=2}^{M-1} \mathbb{E}\left[ \frac{
      (e^{-h\hat{b}_{M-(l-1)}}-e^{-2h\hat{b}_{M-(l-1)}}
      )}{\hat{b}_{M-(l-1)}}  e^{-h\left(
      \sum_{j=1}^{M-l} \hat{b}_j +2 \sum_{j=M-(l-2)}^{M}  \hat{b}_j
      \right) } 
      \right]  + \right.  \notag \\
      & \left. \quad \; \;
 \mathbb{E}\left[ \frac{(e^{-h\hat{b}_{1}}-e^{-2h\hat{b}_{1}})}{\hat{b}_{1}} \right] \prod_{j=2}^M
\mathbb{E}\left[ 
 e^{-2h \hat{b}_j } 
      \right] 
      \right\}  \notag \\
      & = \left\{  \mathbb{E}\left[ \frac{(e^{-h\hat{b} }-e^{-2h\hat{b} })}{\hat{b} }  \right]
    \left(
    \mathbb{E}\left[ e^{-h\hat{b}} 
      \right] \right)^{M-1}  + \sum_{l=2}^{M-1}
  \mathbb{E}\left[ \frac{
      (e^{-h\hat{b}_{M-(l-1)}}-e^{-2h\hat{b}_{M-(l-1)}}
      )}{\hat{b}_{M-(l-1)}} \right]
       \prod_{j=1}^{M-l} \mathbb{E}\left[ e^{-h\hat{b}_j} \right] \prod_{j=M-(l-2)}^{M} \mathbb{E}\left[ e^{-2h\hat{b}_j} \right]  
        + \right. \notag \\
      & \left. \quad \; \;
 \mathbb{E}\left[ \frac{(e^{-h\hat{b} }-e^{-2h\hat{b} })}{\hat{b} } \right] \left(
\mathbb{E}\left[ 
 e^{-2h \hat{b} } 
      \right] \right)^{M-1}
      \right\} \notag \\
      &= \left\{  \mathbb{E}\left[ \frac{(e^{-h\hat{b} }-e^{-2h\hat{b} })}{\hat{b} }  \right]
    \left(
    \mathbb{E}\left[ e^{-h\hat{b}} 
      \right] \right)^{M-1}  + 
      \mathbb{E}\left[ \frac{
      (e^{-h\hat{b}}-e^{-2h\hat{b}}
      )}{\hat{b}} \right]\sum_{l=2}^{M-1} 
       \left( \mathbb{E}\left[ e^{-h\hat{b}} \right] \right)^{M-l} 
       \left( \mathbb{E}\left[ e^{-2h\hat{b}} \right] \right)^{l-1} 
        + \right. \notag \\
      & \left. \quad \; \;
 \mathbb{E}\left[ \frac{(e^{-h\hat{b} }-e^{-2h\hat{b} })}{\hat{b} } \right] \left(
\mathbb{E}\left[ 
 e^{-2h \hat{b} } 
      \right] \right)^{M-1}
      \right\}  \notag \\
      &= \mathbb{E}\left[ \frac{(e^{-h\hat{b} }-e^{-2h\hat{b} })}{\hat{b} }  \right]  \left\{  
    \left(
    \mathbb{E}\left[ e^{-h\hat{b}} 
      \right] \right)^{M-1}  + \frac{\left( \mathbb{E}\left[ e^{-h\hat{b}} \right] \right)^M}{ \mathbb{E}\left[ e^{-2h\hat{b}} \right] }
      \sum_{l=2}^{M-1} 
       \left( \frac{ \mathbb{E}\left[ e^{-2h\hat{b}} \right]}{\mathbb{E}\left[ e^{-h\hat{b}} \right]} \right)^{l} 
        +
\left(
\mathbb{E}\left[ 
 e^{-2h \hat{b} } 
      \right] \right)^{M-1}
      \right\} \notag \\ 
      &=  \mathbb{E}\left[ \frac{(e^{-h\hat{b} }-e^{-2h\hat{b} })}{\hat{b} }  \right]  \left\{  
    \left(
    \mathbb{E}\left[ e^{-h\hat{b}} 
      \right] \right)^{M-1}  + \frac{\left( \mathbb{E}\left[ e^{-h\hat{b}} \right] \right)^M}{ \mathbb{E}\left[ e^{-2h\hat{b}} \right] }
      \frac{\left( \frac{ \mathbb{E}\left[ e^{-2h\hat{b}} \right]}{\mathbb{E}\left[ e^{-h\hat{b}} \right]} \right)^{2} - \left( \frac{ \mathbb{E}\left[ e^{-2h\hat{b}} \right]}{\mathbb{E}\left[ e^{-h\hat{b}} \right]} \right)^{M} }{1- \frac{ \mathbb{E}\left[ e^{-2h\hat{b}} \right]}{\mathbb{E}\left[ e^{-h\hat{b}} \right]}  }
        +
\left(
\mathbb{E}\left[ 
 e^{-2h \hat{b} } 
      \right] \right)^{M-1}
      \right\} \notag \\ 
       &=  \mathbb{E}\left[ \frac{(e^{-h\hat{b} }-e^{-2h\hat{b} })}{\hat{b} }  \right]  \frac{\left( \mathbb{E}\left[ e^{-h\hat{b}} \right] \right)^{M}- \left( \mathbb{E}\left[ e^{-2h\hat{b}} \right] \right)^{M} }{\mathbb{E}\left[ e^{-h\hat{b}} \right] - \mathbb{E}\left[ e^{-2h\hat{b}} \right]}
        \notag \\ 
       &=  \mathbb{E}\left[ \frac{(e^{-h\hat{b} }-e^{-2h\hat{b} })}{\hat{b} }  \right]  \frac{ 
       e^{\frac{Mh}{a-1}} \left( \mathbb{E}\left[ e^{-\frac{h}{a-1}\hat{a}} \right] \right)^{M}- e^{\frac{2Mh}{a-1}} \left( \mathbb{E}\left[ e^{-\frac{2h}{a-1}\hat{a}} \right] \right)^{M} }{
       e^{\frac{h}{a-1}}  \mathbb{E}\left[ e^{-\frac{h}{a-1}\hat{a}} \right] - e^{\frac{2h}{a-1}}  \mathbb{E}\left[ e^{-\frac{2h}{a-1}\hat{a}} \right]}
\end{align}
}

{\allowdisplaybreaks
\begin{align}\label{eq:var_E2}
E_2&= \mathbb{E}\left[
    \frac{\hat{a}_M  }{\hat{b}_M^2 } ( 1- e^{-\hat{b}_M h } )^2 \right] +   \sum_{m=1}^{M-1} 
    \mathbb{E}\left[
    \frac{\hat{a}_m  }{\hat{b}_m^2 } ( 1- e^{-\hat{b}_m h } )^2   e^{-2h\sum_{j=m+1}^M \hat{b}_j } \right] \notag \\
&=
\mathbb{E}\left[
    \frac{\hat{a}_M  }{\hat{b}_M^2 } ( 1- e^{-\hat{b}_M h } )^2 \right] + 
    \sum_{m=1}^{M-1} 
    \left\{ 
    \mathbb{E}\left[
    \frac{\hat{a}_m  }{\hat{b}_m^2 } ( 1- e^{-\hat{b}_m h } )^2  \right]  \mathbb{E}\left[  e^{-2h\sum_{j=m+1}^M \hat{b}_j } \right] \right\} 
    \notag \\
    &=
 \mathbb{E}\left[
    \frac{\hat{a}  }{\hat{b}^2 } ( 1- e^{-\hat{b} h } )^2 \right] \left\{ 
1+ \sum_{m=1}^{M-1} \prod_{j=m+1}^M 
    \mathbb{E}\left[  e^{-2h\hat{b}_j } \right] 
\right\}  \notag  \\ 
&= \mathbb{E}\left[
    \frac{\hat{a}  }{\hat{b}^2 } ( 1- e^{-\hat{b} h } )^2 \right] \left\{ 
1+ \sum_{m=1}^{M-1} \left(
    \mathbb{E}\left[  e^{-2h\hat{b} } \right] \right)^{M-m}
\right\}  \notag \\
&=
\mathbb{E}\left[
    \frac{\hat{a}  }{\hat{b}^2 } ( 1- e^{-\hat{b} h } )^2 \right] \frac{1-\left(\mathbb{E}\left[e^{-2h\hat{b}}\right]\right)^M}{1-\mathbb{E}\left[e^{-2h\hat{b}}\right]} \notag \\
&=
\mathbb{E}\left[ \hat{a}
    \left(\frac{  1- e^{-\hat{b} h }  }{\hat{b} } \right)^2 \right] \frac{1-e^{\frac{2Mh}{a-1}}\left(\mathbb{E}\left[e^{-\frac{2h}{a-1}\hat{a}}\right]\right)^M}{1-e^{\frac{2h}{a-1}}\mathbb{E}\left[e^{-\frac{2h}{a-1}\hat{a}}\right]}
\end{align}
}
As for $E_3$, first separate out the terms for $m\in \{ M-2, M-1\}$ and, subsequently, for $l \in \{1, M-m\}$. Then, note that $$\left(2\sum_{j=m+1}^M \hat{b}_j - \sum_{j=m+1}^{M-(l-1)} \hat{b}_j \right)= \left(\sum_{j=m+1}^M \hat{b}_j+ \sum_{j=M-(l-2)}^M \hat{b}_j \right)= \left(\sum_{j=m+1}^{M-l} \hat{b}_j + 2\sum_{j=M-(l-2)}^M \hat{b}_j + \hat{b}_{M-(l-1)} \right)$$

{\allowdisplaybreaks
\begin{align}
E_3&=  
  \sum_{m=1}^{M-1}
    \mathbb{E}\left[  \frac{\hat{a}_m  }{\hat{b}_m } ( 1- e^{-\hat{b}_m h } )  \left( 
    \sum_{l=1}^{M-m} \frac{(1-e^{-h\hat{b}_{M-(l-1)}})}{\hat{b}_{M-(l-1)}}  e^{-h \left(2\sum_{j=m+1}^M \hat{b}_j - \sum_{j=m+1}^{M-(l-1)} \hat{b}_j \right) }
    \right) \right]
     \notag \\
    & =    \sum_{m=1}^{M-3}
    \mathbb{E}\left[  \frac{\hat{a}_m  }{\hat{b}_m } ( 1- e^{-\hat{b}_m h } )  \left( 
    \sum_{l=1}^{M-m} \frac{(1-e^{-h\hat{b}_{M-(l-1)}})}{\hat{b}_{M-(l-1)}}  e^{-h \left(2\sum_{j=m+1}^M \hat{b}_j - \sum_{j=m+1}^{M-(l-1)} \hat{b}_j \right) }
    \right) \right] + \notag \\
    & \quad \, 
    \mathbb{E}\left[  \frac{\hat{a}_{M-2}  }{\hat{b}_{M-2} } ( 1- e^{-\hat{b}_{M-2} h } )  \left(  \frac{(1-e^{-h\hat{b}_{M}})}{\hat{b}_{M}} e^{-h \left[2\left( \hat{b}_{M-1} + \hat{b}_{M}  \right)  - \left(  \hat{b}_{M-1} + \hat{b}_{M} \right) \right] }  + \right. \right. \notag \\
      &  \left.  \left. \qquad  \qquad \qquad  \qquad \qquad \quad
    \frac{(1-e^{-h\hat{b}_{M-1}})}{\hat{b}_{M-1}} e^{-h \left[2\left( \hat{b}_{M-1} + \hat{b}_{M}  \right)  -  \hat{b}_{M-1}   \right] }
    \right) \right] + \notag \\
    & \quad \, 
    \mathbb{E}\left[  \frac{\hat{a}_{M-1}  }{\hat{b}_{M-1} } ( 1- e^{-\hat{b}_{M-1} h } )   \frac{(1-e^{-h\hat{b}_{M}})}{\hat{b}_{M}} e^{-h \left(2 \hat{b}_{M}    -   \hat{b}_{M}   \right) }
     \right]\notag \\
    & =     \sum_{m=1}^{M-3}
    \mathbb{E}\left[  \frac{\hat{a}_m  }{\hat{b}_m } ( 1- e^{-\hat{b}_m h } )  \frac{(1-e^{-h\hat{b}_{M}})}{\hat{b}_{M}} e^{-h \left(2\sum_{j=m+1}^M \hat{b}_j - \sum_{j=m+1}^{M} \hat{b}_j \right) } \right] + \notag \\ 
    &  \quad  \sum_{m=1}^{M-3}
    \mathbb{E}\left[  \frac{\hat{a}_m  }{\hat{b}_m } ( 1- e^{-\hat{b}_m h } )  \left( 
    \sum_{l=2}^{M-m-1} \frac{(1-e^{-h\hat{b}_{M-(l-1)}})}{\hat{b}_{M-(l-1)}}  e^{-h \left(2\sum_{j=m+1}^M \hat{b}_j - \sum_{j=m+1}^{M-(l-1)} \hat{b}_j \right) }
    \right) \right] + \notag \\
     &  \quad \sum_{m=1}^{M-3}
    \mathbb{E}\left[  \frac{\hat{a}_m  }{\hat{b}_m } ( 1- e^{-\hat{b}_m h } )  \frac{(1-e^{-h\hat{b}_{m+1}})}{\hat{b}_{m+1}} e^{-h \left(2\sum_{j=m+1}^M \hat{b}_j  -\hat{b}_{m+1} \right) } \right]  + \notag 
      \\
    & \quad \, 
    \mathbb{E}\left[  \hat{a}_{M-2}  \frac{ ( 1-e^{-h\hat{b}_{M-2} }  )  }{\hat{b}_{M-2} } \frac{(e^{-h\hat{b}_{M}}-e^{-2h\hat{b}_{M}})}{\hat{b}_{M}}
    e^{-h  \hat{b}_{M-1}  } \right] +   \notag \\
    & \quad \, \mathbb{E}\left[ \hat{a}_{M-2}  \frac{ ( 1-e^{-h\hat{b}_{M-2} }  )  }{\hat{b}_{M-2} }   \frac{(e^{-h\hat{b}_{M-1}}-e^{-2h\hat{b}_{M-1}})}{\hat{b}_{M-1}} e^{-2h \hat{b}_{M} } \right]
    + \notag \\
    & \quad \, 
    \mathbb{E}\left[ \hat{a}_{M-1} \frac{ ( 1-e^{-h\hat{b}_{M-1} } )   }{\hat{b}_{M-1} }   \frac{(e^{-h\hat{b}_{M}}-e^{-2h\hat{b}_{M}})}{\hat{b}_{M}}
     \right]
     \notag   \\
    & =     \sum_{m=1}^{M-3}
    \mathbb{E}\left[  \frac{\hat{a}_m  }{\hat{b}_m } ( 1- e^{-\hat{b}_m h } )  \frac{(1-e^{-h\hat{b}_{M}})}{\hat{b}_{M}} e^{-h \left(\sum_{j=m+1}^{M-1} \hat{b}_j +\hat{b}_M \right) } \right] + \notag \\
     &  \quad  \sum_{m=1}^{M-3}
    \mathbb{E}\left[  \frac{\hat{a}_m  }{\hat{b}_m } ( 1- e^{-\hat{b}_m h } )  \left( 
    \sum_{l=2}^{M-m-1} \frac{(1-e^{-h\hat{b}_{M-(l-1)}})}{\hat{b}_{M-(l-1)}}  e^{-h \left(2\sum_{j=m+1}^M \hat{b}_j - \sum_{j=m+1}^{M-(l-1)} \hat{b}_j \right) }
    \right) \right] + \notag \\
     &  \quad \sum_{m=1}^{M-3}
    \mathbb{E}\left[  \frac{\hat{a}_m  }{\hat{b}_m } ( 1- e^{-\hat{b}_m h } )  \frac{(1-e^{-h\hat{b}_{m+1}})}{\hat{b}_{m+1}} e^{-h \left(2\sum_{j=m+2}^M \hat{b}_j +  \hat{b}_{m+1} \right) } \right]  + \notag \\
       & \quad \, 
    \mathbb{E}\left[  \hat{a}_{M-2}  \frac{ ( 1-e^{-h\hat{b}_{M-2} }  )  }{\hat{b}_{M-2} } \right] \mathbb{E}\left[ \frac{(e^{-h\hat{b}_{M}}-e^{-2h\hat{b}_{M}})}{\hat{b}_{M}} \right] \mathbb{E}\left[
    e^{-h  \hat{b}_{M-1}  } \right] +   \notag \\
    & \quad \, \mathbb{E}\left[ \hat{a}_{M-2}  \frac{ ( 1-e^{-h\hat{b}_{M-2} }  )  }{\hat{b}_{M-2} } \right] \mathbb{E}\left[   \frac{(e^{-h\hat{b}_{M-1}}-e^{-2h\hat{b}_{M-1}})}{\hat{b}_{M-1}} \right] \mathbb{E}\left[ e^{-2h \hat{b}_{M} } \right]
    + \notag \\
    & \quad \, 
    \mathbb{E}\left[ \hat{a}_{M-1} \frac{ (1- e^{-h\hat{b}_{M-1} }  )   }{\hat{b}_{M-1} } \right] \mathbb{E}\left[  \frac{(e^{-h\hat{b}_{M}}-e^{-2h\hat{b}_{M}})}{\hat{b}_{M}}
     \right]
     \notag  \\
      & =     \sum_{m=1}^{M-3}
    \mathbb{E}\left[ \hat{a}_m  \frac{ ( 1-e^{-h\hat{b}_m  }  ) }{\hat{b}_m }   \frac{(e^{-h\hat{b}_{M}}-e^{-2h\hat{b}_{M}})}{\hat{b}_{M}} e^{-h \sum_{j=m+1}^{M-1} \hat{b}_j } \right] + \notag \\
    &  \quad  \sum_{m=1}^{M-3}
    \mathbb{E}\left[  \frac{\hat{a}_m  }{\hat{b}_m } ( 1- e^{-\hat{b}_m h } )  \left( 
    \sum_{l=2}^{M-m-1} \frac{(1-e^{-h\hat{b}_{M-(l-1)}})}{\hat{b}_{M-(l-1)}}  e^{-h \left(2\sum_{j=m+1}^M \hat{b}_j - \sum_{j=m+1}^{M-(l-1)} \hat{b}_j \right) }
    \right) \right] + \notag \\
     &  \quad \sum_{m=1}^{M-3}
    \mathbb{E}\left[ \hat{a}_m \frac{ (1- e^{-h\hat{b}_m  }  )  }{\hat{b}_m }   \frac{(e^{-h\hat{b}_{m+1}}-e^{-2h\hat{b}_{m+1}})}{\hat{b}_{m+1}} e^{-2h \sum_{j=m+2}^M \hat{b}_j   } \right]  + \notag \\
     & \quad \,
    \mathbb{E}\left[ \hat{a} \frac{ ( 1-e^{-h\hat{b} }  )   }{\hat{b} } \right] \mathbb{E}\left[  \frac{(e^{-h\hat{b}}-e^{-2h\hat{b}})}{\hat{b}} \right] \left\{ 1 + \mathbb{E}\left[ e^{-h \hat{b} } \right]+\mathbb{E}\left[ e^{-2h \hat{b} } \right] \right\}  \notag \\
 & = \mathbb{E}\left[ \hat{a} \frac{ (1- e^{-h\hat{b} }  )   }{\hat{b} } \right] \mathbb{E}\left[  \frac{(e^{-h\hat{b}}-e^{-2h\hat{b}})}{\hat{b}} \right] \sum_{m=1}^{M-3} \mathbb{E}\left[e^{-h \sum_{j=m+1}^{M-1} \hat{b}_j } \right]  + \notag \\
    &  \quad  \sum_{m=1}^{M-3}
    \mathbb{E}\left[  \frac{\hat{a}_m  }{\hat{b}_m } ( 1- e^{-\hat{b}_m h } )  \left( 
    \sum_{l=2}^{M-m-1} \frac{(1-e^{-h\hat{b}_{M-(l-1)}})}{\hat{b}_{M-(l-1)}}  e^{-h \left(\sum_{j=m+1}^{M-l} \hat{b}_j + 2\sum_{j=M-(l-2)}^M \hat{b}_j + \hat{b}_{M-(l-1)}  \right) }
    \right) \right] + \notag \\
     &  \quad \,  \mathbb{E}\left[ \hat{a} \frac{ (1- e^{-h\hat{b} }  )   }{\hat{b} } \right] \mathbb{E}\left[  \frac{(e^{-h\hat{b}}-e^{-2h\hat{b}})}{\hat{b}} \right] \sum_{m=1}^{M-3}
     \mathbb{E}\left[ e^{-2h \sum_{j=m+2}^M \hat{b}_j   } \right]  + \notag \\
     & \quad \,
    \mathbb{E}\left[ \hat{a} \frac{ ( 1-e^{-h\hat{b} }  )   }{\hat{b} } \right] \mathbb{E}\left[  \frac{(e^{-h\hat{b}}-e^{-2h\hat{b}})}{\hat{b}} \right] \left\{ 1 + \mathbb{E}\left[ e^{-h \hat{b} } \right] + \mathbb{E}\left[ e^{-2h \hat{b} } \right] \right\} \notag   \\
    & = \mathbb{E}\left[ \hat{a} \frac{ ( 1-e^{-h\hat{b} }  )   }{\hat{b} } \right] \mathbb{E}\left[  \frac{(e^{-h\hat{b}}-e^{-2h\hat{b}})}{\hat{b}} \right]  \sum_{m=1}^{M-3} \prod_{j=m+1}^{M-1} \mathbb{E}\left[e^{-h  \hat{b}_j } \right] + \notag \\
    & \quad 
    \sum_{m=1}^{M-3}
    \mathbb{E}\left[ \hat{a}_m \frac{ (1- e^{-h\hat{b}_m  } ) }{\hat{b}_m }   \left( 
    \sum_{l=2}^{M-m-1} \frac{(e^{-h\hat{b}_{M-(l-1)}}-e^{-2h\hat{b}_{M-(l-1)}})}{\hat{b}_{M-(l-1)}}  e^{-h \left(\sum_{j=m+1}^{M-l} \hat{b}_j + 2\sum_{j=M-(l-2)}^M \hat{b}_j  \right) }
    \right) \right] + \notag \\
    & \quad \,
    \mathbb{E}\left[ \hat{a} \frac{ ( 1-e^{-h\hat{b} }  )   }{\hat{b} } \right] \mathbb{E}\left[  \frac{(e^{-h\hat{b}}-e^{-2h\hat{b}})}{\hat{b}} \right]  \sum_{m=1}^{M-3} \prod_{j=m+2}^{M} \mathbb{E}\left[e^{-2h  \hat{b}_j } \right] +\notag \\
     & \quad \,
    \mathbb{E}\left[ \hat{a} \frac{ (1- e^{-h\hat{b} }  )   }{\hat{b} } \right] \mathbb{E}\left[  \frac{(e^{-h\hat{b}}-e^{-2h\hat{b}})}{\hat{b}} \right] \left\{ 1+ 
     \mathbb{E}\left[ e^{-h \hat{b} } \right]+\mathbb{E}\left[ e^{-2h \hat{b} } \right] 
    \right\} \notag   \\
     & = \mathbb{E}\left[ \hat{a} \frac{ (1- e^{-h\hat{b} }  )   }{\hat{b} } \right] \mathbb{E}\left[  \frac{(e^{-h\hat{b}}-e^{-2h\hat{b}})}{\hat{b}} \right]  \sum_{m=1}^{M-3} \left( \mathbb{E}\left[e^{-h  \hat{b} }  \right] \right)^{M-m-1} + \notag \\
     & \quad 
    \sum_{m=1}^{M-3}
    \mathbb{E}\left[ \hat{a}_m \frac{ ( 1-e^{-h\hat{b}_m  } ) }{\hat{b}_m }  \right] \left(  
    \sum_{l=2}^{M-m-1} \mathbb{E}\left[ \frac{(e^{-h\hat{b}_{M-(l-1)}}-e^{-2h\hat{b}_{M-(l-1)}})}{\hat{b}_{M-(l-1)}} \right]  \mathbb{E}\left[ e^{-h \sum_{j=m+1}^{M-l} \hat{b}_j  }
     \right] \mathbb{E}\left[ e^{-2h \sum_{j=M-(l-2)}^M \hat{b}_j   } \right] \right)+ \notag \\
        & \quad \,
    \mathbb{E}\left[ \hat{a} \frac{ (1-e^{-h\hat{b} }  )   }{\hat{b} } \right] \mathbb{E}\left[  \frac{(e^{-h\hat{b}}-e^{-2h\hat{b}})}{\hat{b}} \right]  \sum_{m=1}^{M-3} \left( \mathbb{E}\left[e^{-2h  \hat{b} } \right] \right)^{M-m-1}+ \notag \\
     & \quad \,
    \mathbb{E}\left[ \hat{a} \frac{ ( 1-e^{-h\hat{b} }  )   }{\hat{b} } \right] \mathbb{E}\left[  \frac{(e^{-h\hat{b}}-e^{-2h\hat{b}})}{\hat{b}} \right] \left\{ 1+ 
     \mathbb{E}\left[ e^{-h \hat{b} } \right]+\mathbb{E}\left[ e^{-2h \hat{b} } \right] 
    \right\} \notag   \\
    & =
    \mathbb{E}\left[ \hat{a} \frac{ ( 1-e^{-h\hat{b} } )   }{\hat{b} } \right] \mathbb{E}\left[  \frac{(e^{-h\hat{b}}-e^{-2h\hat{b}})}{\hat{b}} \right] \left\{ 
    \sum_{m=1}^{M-3} \left( \mathbb{E}\left[e^{-h  \hat{b} } \right] \right)^{M-m-1} + \sum_{m=1}^{M-3} \left( \mathbb{E}\left[e^{-2h  \hat{b} } \right] \right)^{M-m-1}
    \right\}
    + \notag \\
   & \quad \,  \mathbb{E}\left[ \hat{a} \frac{ (1- e^{-h\hat{b} }  )   }{\hat{b} } \right] \mathbb{E}\left[  \frac{(e^{-h\hat{b}}-e^{-2h\hat{b}})}{\hat{b}} \right]  \sum_{m=1}^{M-3} \sum_{l=2}^{M-m-1} \mathbb{E}\left[ e^{-h \sum_{j=m+1}^{M-l} \hat{b}_j  }
     \right] \mathbb{E}\left[ e^{-2h \sum_{j=M-(l-2)}^M \hat{b}_j   } \right] + \notag \\
     & \quad \,
    \mathbb{E}\left[ \hat{a} \frac{ ( 1-e^{-h\hat{b} }  )   }{\hat{b} } \right] \mathbb{E}\left[  \frac{(e^{-h\hat{b}}-e^{-2h\hat{b}})}{\hat{b}} \right] \left\{ 1+ 
     \mathbb{E}\left[ e^{-h \hat{b} } \right]+\mathbb{E}\left[ e^{-2h \hat{b} } \right] 
    \right\} \notag \\
     & =
    \mathbb{E}\left[ \hat{a} \frac{ ( 1-e^{-h\hat{b} }  )   }{\hat{b} } \right] \mathbb{E}\left[  \frac{(e^{-h\hat{b}}-e^{-2h\hat{b}})}{\hat{b}} \right] \left\{ 
    \sum_{m=1}^{M-3} \left( \mathbb{E}\left[e^{-h  \hat{b} } \right] \right)^{M-m-1} + \sum_{m=1}^{M-3} \left( \mathbb{E}\left[e^{-2h  \hat{b} } \right] \right)^{M-m-1}
    \right\}
    + \notag \\
   & \quad \,  \mathbb{E}\left[ \hat{a} \frac{ (1- e^{-h\hat{b} }  )   }{\hat{b} } \right] \mathbb{E}\left[  \frac{(e^{-h\hat{b}}-e^{-2h\hat{b}})}{\hat{b}} \right]  \sum_{m=1}^{M-3} \sum_{l=2}^{M-m-1}
   \prod_{j=m+1}^{M-l} \mathbb{E}\left[ e^{-h\hat{b}_j  }
     \right] \prod_{j=M-(l-2)}^M
    \mathbb{E}\left[ e^{-2h \hat{b}_j   } \right] + \notag \\
     & \quad \,
    \mathbb{E}\left[ \hat{a} \frac{ ( 1-e^{-h\hat{b} }  )   }{\hat{b} } \right] \mathbb{E}\left[  \frac{(e^{-h\hat{b}}-e^{-2h\hat{b}})}{\hat{b}} \right] \left\{ 1+ 
     \mathbb{E}\left[ e^{-h \hat{b} } \right]+\mathbb{E}\left[ e^{-2h \hat{b} } \right] 
    \right\} \notag    \\
     & =
    \mathbb{E}\left[ \hat{a} \frac{ ( 1-e^{-h\hat{b} }  )   }{\hat{b} } \right] \mathbb{E}\left[  \frac{(e^{-h\hat{b}}-e^{-2h\hat{b}})}{\hat{b}} \right] \left\{ 
    \sum_{m=1}^{M-3} \left( \mathbb{E}\left[e^{-h  \hat{b} } \right] \right)^{M-m-1} + \sum_{m=1}^{M-3} \left( \mathbb{E}\left[e^{-2h  \hat{b} } \right] \right)^{M-m-1}
    \right\}
    + \notag \\
   & \quad \,  \mathbb{E}\left[ \hat{a} \frac{ (1- e^{-h\hat{b} }  )   }{\hat{b} } \right] \mathbb{E}\left[  \frac{(e^{-h\hat{b}}-e^{-2h\hat{b}})}{\hat{b}} \right]  \sum_{m=1}^{M-3} \sum_{l=2}^{M-m-1}
   \left( \mathbb{E}\left[ e^{-h\hat{b}  }
     \right] \right)^{M-m-l}  
    \left( \mathbb{E}\left[ e^{-2h \hat{b}   } \right] \right)^{l-1} + \notag \\
     & \quad \,
    \mathbb{E}\left[ \hat{a} \frac{ ( 1-e^{-h\hat{b} }  )   }{\hat{b} } \right] \mathbb{E}\left[  \frac{(e^{-h\hat{b}}-e^{-2h\hat{b}})}{\hat{b}} \right] \left\{ 1+ 
     \mathbb{E}\left[ e^{-h \hat{b} } \right]+\mathbb{E}\left[ e^{-2h \hat{b} } \right] 
    \right\} \notag  \\
         & =
    \mathbb{E}\left[ \hat{a} \frac{ (1- e^{-h\hat{b} }  )   }{\hat{b} } \right] \mathbb{E}\left[  \frac{(e^{-h\hat{b}}-e^{-2h\hat{b}})}{\hat{b}} \right] \left\{ 
    \sum_{m=1}^{M-3} \left( \mathbb{E}\left[e^{-h  \hat{b} } \right] \right)^{M-m-1} + \sum_{m=1}^{M-3} \left( \mathbb{E}\left[e^{-2h  \hat{b} } \right] \right)^{M-m-1}
    \right\}
    + \notag \\
   & \quad \,  \mathbb{E}\left[ \hat{a} \frac{ (1- e^{-h\hat{b} }  )   }{\hat{b} } \right] \mathbb{E}\left[  \frac{(e^{-h\hat{b}}-e^{-2h\hat{b}})}{\hat{b}} \right] 
   \left\{ \frac{ \mathbb{E}\left[ e^{-2h \hat{b} } \right] }{\mathbb{E}\left[ e^{-h \hat{b} } \right]-\mathbb{E}\left[ e^{-2h \hat{b} } \right]}  \sum_{m=1}^{M-3} \left( \mathbb{E}\left[e^{-h  \hat{b} } \right] \right)^{M-m-1} \right\}
  -\notag \\
   & \quad \,  \mathbb{E}\left[ \hat{a} \frac{ (1- e^{-h\hat{b} }  )   }{\hat{b} } \right] \mathbb{E}\left[  \frac{(e^{-h\hat{b}}-e^{-2h\hat{b}})}{\hat{b}} \right] 
   \left\{ \frac{ \mathbb{E}\left[ e^{-h \hat{b} } \right] }{\mathbb{E}\left[ e^{-h \hat{b} } \right]-\mathbb{E}\left[ e^{-2h \hat{b} } \right]}  \sum_{m=1}^{M-3} \left( \mathbb{E}\left[e^{-2h  \hat{b} } \right] \right)^{M-m-1} \right\}
  + \notag \\
     & \quad \,
    \mathbb{E}\left[ \hat{a} \frac{ (1- e^{-h\hat{b} }  )   }{\hat{b} } \right] \mathbb{E}\left[  \frac{(e^{-h\hat{b}}-e^{-2h\hat{b}})}{\hat{b}} \right] \left\{ 1+ 
     \mathbb{E}\left[ e^{-h \hat{b} } \right]+\mathbb{E}\left[ e^{-2h \hat{b} } \right] 
    \right\} \notag  \\
      & =
    \mathbb{E}\left[ \hat{a} \frac{ ( 1-e^{-h\hat{b} }  )   }{\hat{b} } \right] \mathbb{E}\left[  \frac{(e^{-h\hat{b}}-e^{-2h\hat{b}})}{\hat{b}} \right]  \frac{\mathbb{E}\left[e^{-h  \hat{b} } \right]}{\mathbb{E}\left[e^{-h  \hat{b} } \right]-\mathbb{E}\left[e^{-2h  \hat{b} } \right]} 
    \sum_{m=1}^{M-3} \left( \mathbb{E}\left[e^{-h  \hat{b} } \right] \right)^{M-m-1} -
     \notag \\ & \quad \,
    \mathbb{E}\left[ \hat{a} \frac{ (1- e^{-h\hat{b} }  )   }{\hat{b} } \right] \mathbb{E}\left[  \frac{(e^{-h\hat{b}}-e^{-2h\hat{b}})}{\hat{b}} \right] \frac{\mathbb{E}\left[e^{-2h  \hat{b} } \right]}{\mathbb{E}\left[e^{-h  \hat{b} } \right]-\mathbb{E}\left[e^{-2h  \hat{b} } \right]}  \sum_{m=1}^{M-3} \left( \mathbb{E}\left[e^{-2h  \hat{b} } \right] \right)^{M-m-1} +\notag \\
    & \quad \,
    \mathbb{E}\left[ \hat{a} \frac{ (1- e^{-h\hat{b} }  )   }{\hat{b} } \right] \mathbb{E}\left[  \frac{(e^{-h\hat{b}}-e^{-2h\hat{b}})}{\hat{b}} \right] \left\{ 1+ 
     \mathbb{E}\left[ e^{-h \hat{b} } \right]+\mathbb{E}\left[ e^{-2h \hat{b} } \right] 
    \right\} \notag  \\
        & =
    \mathbb{E}\left[ \hat{a} \frac{ (1- e^{-h\hat{b} }  )   }{\hat{b} } \right] \mathbb{E}\left[  \frac{(e^{-h\hat{b}}-e^{-2h\hat{b}})}{\hat{b}} \right]  \frac{\mathbb{E}\left[e^{-h  \hat{b} } \right]}{\mathbb{E}\left[e^{-h  \hat{b} } \right]-\mathbb{E}\left[e^{-2h  \hat{b} } \right]} 
    \frac{\left( \mathbb{E}\left[e^{-h  \hat{b} } \right] \right)^{M-1}-\left( \mathbb{E}\left[e^{-h  \hat{b} } \right] \right)^{2}}{ \mathbb{E}\left[e^{-h  \hat{b} } \right] -1}  -
     \notag \\ & \quad \,
    \mathbb{E}\left[ \hat{a} \frac{ (1- e^{-h\hat{b} }  )   }{\hat{b} } \right] \mathbb{E}\left[  \frac{(e^{-h\hat{b}}-e^{-2h\hat{b}})}{\hat{b}} \right] \frac{\mathbb{E}\left[e^{-2h  \hat{b} } \right]}{\mathbb{E}\left[e^{-h  \hat{b} } \right]-\mathbb{E}\left[e^{-2h  \hat{b} } \right]}  \frac{\left( \mathbb{E}\left[e^{-2h  \hat{b} } \right] \right)^{M-1}-\left( \mathbb{E}\left[e^{-2h  \hat{b} } \right] \right)^{2}}{ \mathbb{E}\left[e^{-2h  \hat{b} } \right] -1}  +\notag \\
    & \quad \,
    \mathbb{E}\left[ \hat{a} \frac{ (1- e^{-h\hat{b} }  )   }{\hat{b} } \right] \mathbb{E}\left[  \frac{(e^{-h\hat{b}}-e^{-2h\hat{b}})}{\hat{b}} \right] \left\{ 1+ 
     \mathbb{E}\left[ e^{-h \hat{b} } \right]+\mathbb{E}\left[ e^{-2h \hat{b} } \right] 
    \right\} \notag \\
         & =
    \mathbb{E}\left[ \hat{a} \frac{ ( 1-e^{-h\hat{b} }  )   }{\hat{b} } \right] \mathbb{E}\left[  \frac{(e^{-h\hat{b}}-e^{-2h\hat{b}})}{\hat{b}} \right] \times \notag \\
    & \quad 
    \left\{ 1+ 
     \mathbb{E}\left[ e^{-h \hat{b} } \right]+\mathbb{E}\left[ e^{-2h \hat{b} } \right] +  \frac{\left( \mathbb{E}\left[e^{-h  \hat{b} } \right] \right)^{M}-\left( \mathbb{E}\left[e^{-h  \hat{b} } \right] \right)^{3}}{ \left(
    \mathbb{E}\left[e^{-h  \hat{b} } \right]-\mathbb{E}\left[e^{-2h  \hat{b} } \right] 
    \right) \left( \mathbb{E}\left[e^{-h  \hat{b} } \right] -1 \right)} - \frac{\left( \mathbb{E}\left[e^{-2h  \hat{b} } \right] \right)^{M}-\left( \mathbb{E}\left[e^{-2h  \hat{b} } \right] \right)^{3}}{ \left(\mathbb{E}\left[e^{-h  \hat{b} } \right]-\mathbb{E}\left[e^{-2h  \hat{b} } \right] \right) \left(\mathbb{E}\left[e^{-2h  \hat{b} } \right] -1\right)} 
    \right\}
    \notag \\
         & =
    \mathbb{E}\left[ \hat{a} \frac{ ( 1-e^{-h\hat{b} }  )   }{\hat{b} } \right] \mathbb{E}\left[  \frac{(e^{-h\hat{b}}-e^{-2h\hat{b}})}{\hat{b}} \right] \times \notag \\
    & \; \; \;
    \left\{
     e^{\frac{h}{a-1}}\mathbb{E}\left[ e^{-\frac{h}{a-1} \hat{a} } \right]+  e^{\frac{2h}{a-1}} \mathbb{E}\left[ e^{-\frac{2h}{a-1} \hat{a} } \right] + 
    \frac{ e^{\frac{Mh}{a-1}}\left( \mathbb{E}\left[e^{-\frac{h}{a-1}  \hat{a} } \right] \right)^{M}-  e^{\frac{3h}{a-1}} \left( \mathbb{E}\left[e^{-\frac{h}{a-1}  \hat{a} } \right] \right)^{3}}{ \left(
     e^{\frac{h}{a-1}}\mathbb{E}\left[e^{-\frac{h}{a-1}  \hat{a} } \right]- e^{\frac{2h}{a-1}}\mathbb{E}\left[e^{-\frac{2h}{a-1}  \hat{a} } \right] 
    \right) \left(  e^{\frac{h}{a-1}} \mathbb{E}\left[e^{-\frac{h}{a-1}  \hat{a} } \right] -1 \right)} -
          \right. 
    \notag \\ &
    \left. \quad \;
    \frac{  e^{\frac{2Mh}{a-1}} \left(\mathbb{E}\left[e^{-\frac{2h}{a-1}  \hat{a} } \right] \right)^{M}-  e^{\frac{6h}{a-1}} \left( \mathbb{E}\left[e^{-\frac{2h}{a-1}  \hat{a} } \right] \right)^{3}}{ \left( e^{\frac{h}{a-1}} \mathbb{E}\left[e^{-\frac{h}{a-1}  \hat{a} } \right]- e^{\frac{2h}{a-1}} \mathbb{E}\left[e^{-\frac{2h}{a-1}  \hat{a} } \right] \right) \left(e^{\frac{2h}{a-1}}\mathbb{E}\left[e^{-\frac{2h}{a-1}  \hat{a} } \right] -1\right)} +1 
    \right\}
\end{align}
}
Armed with these results, \ref{eq:var_second_exp} now becomes:
{\allowdisplaybreaks
\begin{align}
    \mathbb{V}\text{ar}\left[\hat{\theta}_M  \right] & = 2\theta_0 \mathbb{E}\left[ \frac{(e^{-h\hat{b} }-e^{-2h\hat{b} })}{\hat{b} }  \right]  \frac{ 
       e^{\frac{Mh}{a-1}} \left( \mathbb{E}\left[ e^{-\frac{h}{a-1}\hat{a}} \right] \right)^{M}- e^{\frac{2Mh}{a-1}} \left( \mathbb{E}\left[ e^{-\frac{2h}{a-1}\hat{a}} \right] \right)^{M} }{
       e^{\frac{h}{a-1}}  \mathbb{E}\left[ e^{-\frac{h}{a-1}\hat{a}} \right] - e^{\frac{2h}{a-1}}  \mathbb{E}\left[ e^{-\frac{2h}{a-1}\hat{a}} \right]} + \notag \\
    & \quad \mathbb{E}\left[ \hat{a}
    \left(\frac{  1- e^{-\hat{b} h }  }{\hat{b} } \right)^2 \right] \frac{1-e^{\frac{2Mh}{a-1}}\left(\mathbb{E}\left[e^{-\frac{2h}{a-1}\hat{a}}\right]\right)^M}{1-e^{\frac{2h}{a-1}}\mathbb{E}\left[e^{-\frac{2h}{a-1}\hat{a}}\right]} + \notag \\
    & \quad 
        2  \mathbb{E}\left[ \hat{a} \frac{ ( 1-e^{-h\hat{b} }  )   }{\hat{b} } \right] \mathbb{E}\left[  \frac{(e^{-h\hat{b}}-e^{-2h\hat{b}})}{\hat{b}} \right] \times \notag \\
    & \; \; \;
  \left\{
     1 +
     e^{\frac{h}{a-1}}\mathbb{E}\left[ e^{-\frac{h}{a-1} \hat{a} } \right]+  e^{\frac{2h}{a-1}} \mathbb{E}\left[ e^{-\frac{2h}{a-1} \hat{a} } \right] + 
    \frac{ e^{\frac{Mh}{a-1}}\left( \mathbb{E}\left[e^{-\frac{h}{a-1}  \hat{a} } \right] \right)^{M}-  e^{\frac{3h}{a-1}} \left( \mathbb{E}\left[e^{-\frac{h}{a-1}  \hat{a} } \right] \right)^{3}}{ \left(
     e^{\frac{h}{a-1}}\mathbb{E}\left[e^{-\frac{h}{a-1}  \hat{a} } \right]- e^{\frac{2h}{a-1}}\mathbb{E}\left[e^{-\frac{2h}{a-1}  \hat{a} } \right] 
    \right) \left(  e^{\frac{h}{a-1}} \mathbb{E}\left[e^{-\frac{h}{a-1}  \hat{a} } \right] -1 \right)} -
          \right. 
    \notag \\ &
    \left. \quad \; 
    \frac{  e^{\frac{2Mh}{a-1}} \left(\mathbb{E}\left[e^{-\frac{2h}{a-1}  \hat{a} } \right] \right)^{M}-  e^{\frac{6h}{a-1}} \left( \mathbb{E}\left[e^{-\frac{2h}{a-1}  \hat{a} } \right] \right)^{3}}{ \left( e^{\frac{h}{a-1}} \mathbb{E}\left[e^{-\frac{h}{a-1}  \hat{a} } \right]- e^{\frac{2h}{a-1}} \mathbb{E}\left[e^{-\frac{2h}{a-1}  \hat{a} } \right] \right) \left(e^{\frac{2h}{a-1}}\mathbb{E}\left[e^{-\frac{2h}{a-1}  \hat{a} } \right] -1\right)} 
    \right\}  
    \notag \\ 
    &=
2\theta_0 \mathbb{E}\left[ \frac{(e^{-h\hat{b} }-e^{-2h\hat{b} })}{\hat{b} }  \right]  \frac{ 
       e^{\frac{Mh}{a-1}} \left( 
       \mathcal{M}_{\hat{a}}\left( -\frac{h}{a-1}\right)
       \right)^{M}- e^{\frac{2Mh}{a-1}} \left( 
       \mathcal{M}_{\hat{a}}\left( -\frac{2h}{a-1}\right)
       \right)^{M} 
       }{
       e^{\frac{h}{a-1}}  
       \mathcal{M}_{\hat{a}}\left( -\frac{h}{a-1}\right) - e^{\frac{2h}{a-1}}  \mathcal{M}_{\hat{a}}\left( -\frac{2h}{a-1}\right)} + \notag \\
    & \quad \quad  \mathbb{E}\left[ \hat{a}
   \left( \frac{  1- e^{-\hat{b} h }   }{\hat{b} } \right)^2  \right] \frac{1-e^{\frac{2Mh}{a-1}}\left(
    \mathcal{M}_{\hat{a}}\left( -\frac{2h}{a-1}\right)
    \right)^M}{1-e^{\frac{2h}{a-1}}
    \mathcal{M}_{\hat{a}}\left( -\frac{2h}{a-1}\right)
    } + \notag \\
    & \quad 
       \quad 
        2  \mathbb{E}\left[ \hat{a} \frac{ ( 1-e^{-h\hat{b} }  )   }{\hat{b} } \right] \mathbb{E}\left[  \frac{(e^{-h\hat{b}}-e^{-2h\hat{b}})}{\hat{b}} \right] \times \notag \\
    & \; \; \;
  \left\{      \vphantom{\frac{\mathbb{E}\left[ e^{-2h \hat{b} } \right]}{\mathbb{E}\left[ e^{-2h \hat{b} } \right]}}
     1 +
     e^{\frac{h}{a-1}} \mathcal{M}_{\hat{a}}\left( -\frac{h}{a-1}\right) +  e^{\frac{2h}{a-1}} \mathcal{M}_{\hat{a}}\left( -\frac{2h}{a-1}\right) +  \right. \notag  \\ 
     & \left. \quad \;
    \frac{ e^{\frac{Mh}{a-1}}\left( \mathcal{M}_{\hat{a}}\left( -\frac{h}{a-1}\right) \right)^{M}-  e^{\frac{3h}{a-1}} \left( \mathcal{M}_{\hat{a}}\left( -\frac{h}{a-1}\right) \right)^{3}}{ \left(
     e^{\frac{h}{a-1}}
     \mathcal{M}_{\hat{a}}\left( -\frac{h}{a-1}\right)
     - e^{\frac{2h}{a-1}}
    \mathcal{M}_{\hat{a}}\left( -\frac{2h}{a-1}\right) 
    \right) \left(  e^{\frac{h}{a-1}} \mathcal{M}_{\hat{a}}\left( -\frac{h}{a-1}\right) -1 \right)} -
          \right. 
    \notag \\ &
    \left. \quad \; 
    \frac{  e^{\frac{2Mh}{a-1}} \left(\mathcal{M}_{\hat{a}}\left( -\frac{2h}{a-1}\right) \right)^{M}-  e^{\frac{6h}{a-1}} \left( \mathcal{M}_{\hat{a}}\left( -\frac{2h}{a-1}\right) \right)^{3}}{ \left( e^{\frac{h}{a-1}} \mathcal{M}_{\hat{a}}\left( -\frac{h}{a-1}\right)- e^{\frac{2h}{a-1}} \mathcal{M}_{\hat{a}}\left( -\frac{2h}{a-1}\right) \right) \left(e^{\frac{2h}{a-1}}\mathcal{M}_{\hat{a}}\left( -\frac{2h}{a-1}\right) -1\right)} 
    \right\}  
\end{align}
}
where, as before, $\mathcal{M}_{\hat{a}}\left( t \right)= \mathbb{E}\left[ e^{t\hat{a} }\right]$ is the moment-generating function of $\hat{a}$, $t=-\frac{h}{a-1}$ and $\mathbb{E}\left[\phi\left( \hat{a}_m \right)\right]=\mathbb{E}\left[\phi\left( \hat{a} \right)\right]$ $\forall m=1, \ldots, M$ for  non-linear functions $\phi\left( \cdot \right)$ as in \eqref{eq:mean_scir_cv_2}-\eqref{eq:variance_scir_cv_2}.
\end{proof}

\section{Experiments}\label{app:app4}
\subsection*{Latent Dirichlet Allocation}
The latent Dirichlet allocation 
model \citep{LDA_Blei}  consists of
$K$ topics each with its distribution $\omega_k$ over the $W$ words in the vocabulary, drawn from a symmetric Dirichlet prior with hyper-parameter $\beta$. A document $w_d$ is modeled as a mixture of topics, with mixing weight $\eta_d$, drawn from a symmetric Dirichlet prior with hyper-parameter $\alpha$. The model is a
generative process where documents are produced as a set of words by drawing a topic assignment $z_{di}\stackrel{\text {iid}}{\sim} \eta_d$ for each word $w_{di}$ in document $w_d$ and then drawing the word from the corresponding topic $\omega_{z_{di}}$.  
Conditional on $\omega$, documents are i.i.d.:
\begin{align}\label{eq:lda_joint}
    p\left( w, z, \omega \mid \alpha, \beta \right) = p\left( \omega \mid \beta \right) \prod_{d=1}^D p\left( w_d, z_d \mid \alpha, \omega  \right)
\end{align}
where $p\left( w_d, z_d \mid \alpha, \omega  \right)=\prod_{k=1}^K \frac{\Gamma(\alpha + n_{dk \cdot})}{\Gamma(\alpha)} \prod_{w=1}^W \omega_{kw}^{n_{dkw}}$ and, as in \cite{vb_lda}, $n_{dkw}=\sum_{i=1}^{N_d} \delta(w_{di}=w, z_{di}=k)$ and $\cdot$ denotes summation over the
corresponding index. 
To apply the control variate-based stochastic Cox-Ingersoll-Ross Algorithm, it suffices to recognize that the latent Dirichlet allocation posterior can be expressed as a transformation of independent gamma random variables. For each
of the $K$ topics $\omega_k$, we introduce a $W$-dimensional parameter $\theta_{kw}$ for the unnormalised categorical probability. Then $\omega_{kw}=\frac{\theta_{kw}}{\sum_{w=1}^W \theta_{kw}}$. The algorithm runs on mini-batches of documents: at time $t$ it receives a mini-batch of documents
indexed by $D_t$, drawn at random from the full corpus $D$. The stochastic gradient of the log posterior
of $\theta$ on $D_t$ as in \eqref{eq:cv_est} can be derived from the joint distribution in \eqref{eq:lda_joint} by Fisher's identity \citep[see][Appendix D]{Douc2013} and is:
\begin{align*}
    \frac{\partial}{\partial \theta_{kw}} \log \pi\left( \theta \mid w, \alpha, \beta \right) = 
    \frac{1 }{\theta_{kw}} \left( \beta +  \frac{|D|}{|D_t|} \sum_{d \in D_t} \hat{z}_{dkw} - 1 \right)  - \frac{\beta + \frac{|D|}{|D_t|} \sum_{d \in D_t} \hat{z}_{dkw} -1}{\beta + \sum_{d \in D} \hat{z}_{dkw} -1}
\end{align*}
where 
$\hat{z}_{dkw}=
  \mathbb{E}_{z_{dkw} \mid w_d, \theta, \alpha} \left[ 
  n_{dkw} - \omega_{kw} n_{dk\cdot}
  \right]
$.
The observed counts $\sum_{i \in S} z_i$ in \eqref{eq:scir_cv}
have now been replaced with the expectation of the latent topic assignment counts $n_{dkw}$.
As in \cite{SGRLD_Patterson_Teh}, 
to calculate this expectation we use Gibbs sampling on the topic assignments in each document separately, using the conditional distributions
\begin{align}\label{eq:gibb_topic_ass}
    p\left( z_{di} = k \mid w_d, \theta, \alpha \right)= \frac{ \left( \alpha + n_{dk\cdot}^{\backslash i } \right) \theta_{kw_{di}}}{\sum_k \left( \alpha + n_{dk\cdot}^{\backslash i } \right)\theta_{kw_{di}}} 
\end{align}
where $\backslash i$ represents a count excluding the topic assignment variable we are updating.\\
Performance is evaluated by measuring the predictive ability of the trained model on a held-out test set. 
A metric frequently used for this purpose is perplexity, the exponentiated
cross entropy between the trained model probability distribution and the empirical distribution of the test data. The perplexity is monotonically decreasing in the likelihood of the test data and is algebraicly equivalent to the inverse of the geometric mean per-word likelihood. A
lower perplexity score indicates better generalization performance.
More formally, for a held-out document $w_d$ and a training set $\mathcal{W}$, the perplexity is given by 
\begin{align}\label{eq:perp}
    \text{perp}\left( w_d \mid \mathcal{W}, \alpha, \beta \right)=\exp{ \left\{ -\frac{ \sum_{i=1}^{n_{d \cdot \cdot}} \log p\left( w_{di} \mid \mathcal{W}, \alpha, \beta \right) }{n_{d \cdot \cdot} } \right\} }
\end{align}
We use a document completion approach \citep{perplex}, partitioning the test document $w_d$ into two disjoint sets of words $w_d^{\text{train}}$ and $w_d^{\text{test}}$ and using $w_d^{\text{train}}$ to estimate $\eta_d$ for the test document and then calculating the perplexity on $w_d^{\text{test}}$ using this estimate, that is $\log p\left( w_{di} \mid \mathcal{W}, \alpha, \beta \right)$ in \eqref{eq:perp} is replace by
\begin{align*}
     \log p\left( w_{di} \mid w_d^{\text{train}}, \mathcal{W}, \alpha, \beta \right)= \mathbb{E}_{\omega \mid \mathcal{W}, \alpha, \beta } \left[ \mathbb{E}_{z_d^{\text{train}} \mid \omega, \alpha } \left[ \sum_k \hat{\eta}_{dk} \omega_{kw_{di}} \right] \right]
\end{align*}
where 
 $\hat{\eta}_{dk} = p\left( z_{di}^\text{test}=k \mid  z_d^{\text{train}}, \alpha  \right)=\frac{n_{dk\cdot}^{\text{train}} + \alpha}{n_{d\cdot\cdot}^{\text{train}} + K \alpha}$.
We estimate these expectations using the samples we obtain for $\omega$ from the Markov chain produced
by the control variate-based stochastic Cox-Ingersoll-Ross Algorithm and samples for $z_d^{\text{train}}$ produced by Gibbs sampling the topic assignments on $w_d^{\text{train}}$.\\
We apply the control variate-based stochastic Cox-Ingersoll-Ross process, the stochastic Cox-Ingersoll-Ross process \citep{Baker2018LargeScaleSS} and the stochastic Gradient Riemannian Langevin dynamics \citep{SGRLD_Patterson_Teh} to sample from the latent Dirichlet allocation model applied to a dataset of scraped Wikipedia documents.
The vocabulary used is 
as in \cite{OnLearLDA}; it is taken from the top $10\,000$ words in Project
Gutenburg texts, excluding all words of less than three characters. This results in a vocabulary size $W$ of approximately $8\,000$ words. 
In total $|D|=50\,000$ documents from Wikipedia were used, in minibatches
of $|D_t|=50$ documents each. The perplexities were estimated
on a separate holdout set of $1\,000$ documents, split $90/10$ training/test.  
Similar to \cite{Baker2018LargeScaleSS, SGRLD_Patterson_Teh}, for all methods, we use a decreasing stepsize
scheme of the form $h_t=h\left[ 1+t/\tau\right]^{-\kappa}$. Details on the hyperparameters used are given in \autoref{tab1}. R code for the experiment is available online at the following link: \url{https://github.com/FrancescoBarile/SCIR_CV}.

\begin{table}
\centering
Hyperparameters for the latent Dirichlet allocation experiment
\begin{tabular}{crrrcrrrrc}
		\hline
		Method & $h$ & $\tau$ & $\kappa$ & $\alpha$ & $\beta$ & $K$ & $|D_t|$ & $\ell$ & Gibbs Samples \\ 
		\hline
		Baker et al. (2018) & $1$ & $1000$ & $3.32$ & $1.1$ & $0.1$ & $50$ & $50$ & - & $200$\\ 
		Patterson \& Teh (2013) & $1$ & $1000$ & $3.32$ &  $1.1$ & $0.1$ & $50$ & $50$ & - & $200$\\ 
		Our proposed method & $1$ & $1000$ & $3.32$ & $1.1$ & $0.1$ & $50$ & $50$ & $5$ & $200$\\ 
		\hline
\end{tabular}\caption{
    Gibbs Samples, iterations to estimate topic assignment as in \ref{eq:gibb_topic_ass}. Every $\ell$ iterations estimate the posterior parameter $a$ based on $1\,000$ documents.
}\label{tab1}
\end{table}

\end{bibunit}

\end{document}